\documentclass[%
reprint, 
superscriptaddress,
 amsmath,amssymb,
 aps, physrev,
pra,
]{revtex4-2}
\usepackage{float}
\usepackage[english]{babel}
\makeatletter
\renewcommand{\selectlanguage}[1]{}%
\makeatother
\usepackage{multirow}
\usepackage{amssymb}   
\usepackage{braket}
\usepackage{graphicx}
\usepackage{dcolumn}
\usepackage{bm}
\usepackage[markup=underlined]{changes}

\begin{document}

\preprint{APS/123-QED}

\title[NbTiN Nanowire Resonators and Spin–Photon Coupling Prospects with Electrons on Solid Neon]{NbTiN Nanowire Resonators and Spin–Photon Coupling Prospects with Electrons on Solid Neon}
\author{Y. Tian}
\affiliation{RIKEN Center for Quantum Computing, 2-1 Hirosawa, Wako, Saitama, 351-0198, Japan}
\author{I. Grytsenko}
\affiliation{RIKEN Center for Quantum Computing, 2-1 Hirosawa, Wako, Saitama, 351-0198, Japan}
\author{A. Jennings}
\affiliation{RIKEN Center for Quantum Computing, 2-1 Hirosawa, Wako, Saitama, 351-0198, Japan}
\author{J. Wang}
\affiliation{RIKEN Center for Quantum Computing, 2-1 Hirosawa, Wako, Saitama, 351-0198, Japan}
\author{H. Ikegami}
\affiliation{RIKEN Center for Quantum Computing, 2-1 Hirosawa, Wako, Saitama, 351-0198, Japan}
\author{X. Zhou}%
\affiliation{RIKEN Center for Quantum Computing, 2-1 Hirosawa, Wako, Saitama, 351-0198, Japan}
\affiliation{Department of Mechanical Engineering, FAMU-FSU College of Engineering, Florida State University, Tallahassee, Florida 32310, USA}
\author{S. Tamate}%
\affiliation{RIKEN Center for Quantum Computing, 2-1 Hirosawa, Wako, Saitama, 351-0198, Japan}
\author{H. Terai}%
\affiliation{Advanced ICT Research Institute, National Institute of Information and Communications Technology (NICT), Kobe, 651-2492, Japan}
\author{H. Kutsuma}%
\affiliation{Graduate School of Engineering, Tohoku University, Sendai 980-8579, Japan}
\author{D. Jin}\affiliation{Department of Physics and Astronomy, University of Notre Dame, Notre Dame, Indiana 46556, USA}
\author{M. Benito}%
\affiliation{Institute of Physics, University of Augsburg, Augsburg, 86159, Germany}
\affiliation{Center for Advanced Analytics and Predictive Sciences, University of Augsburg, 86135 Augsburg, Germany}
\author{E. Kawakami}
\email{e2006k@gmail.com}
\affiliation{RIKEN Center for Quantum Computing, 2-1 Hirosawa, Wako, Saitama, 351-0198, Japan}
\affiliation{RIKEN Cluster for Pioneering Research, 2-1 Hirosawa, Wako, Saitama, 351-0198, Japan.}
\date{\today}

\begin{abstract}
Electrons floating on a solid neon exhibit long charge coherence times, making them attractive for hybrid quantum systems. When combined with high-quality, high-impedance superconducting resonators and a local magnetic field gradient, this platform enables strong charge--photon and spin--charge coupling—key ingredients for scalable spin qubit architectures. 
In this work, we demonstrate that NbTiN nanowire resonators maintain high quality factors around $10^5$ after depositing solid neon onto the resonators and subsequently loading electrons onto the neon surface, validating their suitability for electrons-on-neon platforms. Building on these experimental results, we theoretically analyze micromagnet designs and coupling strategies that can enable spin–photon interactions in this platform. Our analysis outlines performance targets for next-generation devices, showing that, at the charge sweet spot, spin qubit gate fidelities exceeding 99.99\% for single-qubit operations and 99.9\% for two-qubit operations are achievable with natural neon.
\end{abstract}

\maketitle


\section{\label{sec:Intro}Introduction}

Using cavity quantum electrodynamics (cQED), the quantum states of an electron can be coupled to a microwave photon in a superconducting resonator. This approach has been successfully implemented using electrons in semiconductors, for both charge~\cite{Petersson2012,Mi2017-sp,Stockklauser2017-ao} and spin~\cite{Landig2018-wt,Samkharadze2016-xh,Mi2018-co} degrees of freedom. Recent experiments have demonstrated coupling between distant qubits~\cite{Borjans2020-do,Bottcher2022-dp,Harvey-Collard2022-jh} and, more recently, realized remote two-qubit gate operations~\cite{Dijkema2025}, contributing to the development of scalable electron spin qubit systems.

Recently, cQED experiments have been demonstrated using electrons on solid neon~\cite{Zhou2022-nk}. The development of an electron charge qubit on solid neon is a significant breakthrough, reaching charge coherence times of about $100\,\mu\mathrm{s}$~\cite{Zhou2022-nk,Zhou2023-iw}. Recent theoretical studies suggest that at temperatures around $\sim10\,\mathrm{mK}$, decoherence of the electron's charge degree of freedom is primarily limited by acoustic phonons in solid neon~\cite{Li2025-em}. However, the coherence time expected from this mechanism is longer than the measured value, which indicates that the observed decoherence may instead be dominated by stray electrons near the qubit~\cite{li2025electron}. Therefore, controlling the electron density on solid neon is a key step toward achieving qubits with longer coherence times.

Even longer coherence times, potentially reaching the order of $1\,\mathrm{s}$, are expected for the spin states of an electron on solid neon~\cite{Chen2022-on},  making them attractive for spin-photon coupling schemes. Such coupling, mediated by charge-photon interaction, has been proposed~\cite{Benito2017-ok,Benito2019-al} and experimentally demonstrated with electrons in semiconductors~\cite{Mi2018-co,Samkharadze2018,Dijkema2025}. In this framework, maintaining long coherence times for both spin and charge states is crucial for achieving high spin qubit gate fidelities. 

By combining cQED technology with these extended coherence times for both the charge and spin states of electrons on solid neon, electron qubit systems can be scaled into large networks, while simultaneously simplifying the complexity of control required for quantum error correction~\cite{Fowler2012}. Along this line, the primary goal is to achieve strong coupling between the spin state of a single electron on solid neon and a microwave photon (spin-photon strong coupling). In order to realize spin-photon strong coupling, spin-charge interaction is required. The charge state can couple to a photon in a superconducting resonator, enabling spin-photon coupling via the spin-charge interaction. This spin-charge interaction can be artificially introduced using locally placed micro-ferromagnets~\cite{Tokura2006,Jennings2024-sb,Kawakami2023-vf,Schuster2010}. An external magnetic field is applied to magnetize the ferromagnet and induce a finite Zeeman splitting. 

As a first step toward realizing spin qubits using electrons on neon and a nanowire resonator, we fabricated NbTiN superconducting nanowire resonators without integrating micromagnets, and carried out foundational experiments. In these experiments, we deposited a thin layer of neon followed by electrons onto the resonator, gradually increased the electron density, and monitored the resulting resonance peak shift. Importantly, depositing neon and electrons did not degrade the internal quality factor of the resonators. Encouraged by these experimental results, we proceeded to a theoretical analysis of the next stage toward implementing spin qubits, using the experimentally obtained values as input. We investigated the optimized geometry for micromagnets required to induce local magnetic field gradients, and demonstrated that cooperativity between the resonator photons and the spin state can reach \(\gtrsim 10^7\). We theoretically identified optimal conditions that enable high-fidelity single- and two-qubit gate operations for spin qubits, and found that fidelities of 99.99\% and 99.9\%, respectively, are potentially achievable even when using natural neon.

\section{\label{sec:NWresonator}nanowire resonator}

Since applying a magnetic field is necessary, we fabricated resonators using a $20\,\text{nm}$-thick NbTiN film sputtered onto high-resistivity silicon to enhance their resilience to magnetic fields. In addition, NbTiN has high kinetic inductance, which, in turn, strengthens the coupling between the resonator's photons and the charge state~\cite{Samkharadze2016-xh}. The critical temperature of the NbTiN film was measured to be $10.7\,\text{K}$. Achieving strong coupling between the resonator's photons and the charge state is crucial for obtaining high spin-photon cooperativity, which is essential for realizing high spin qubit gate fidelities~\cite{Jennings2024-sb,Dijkema2025}. To further improve performance, a nanowire-shaped resonator was chosen to suppress the generation of vortices that could disrupt superconductivity~\cite{Samkharadze2016-xh,Kroll2019-eo}.

\begin{figure}
    \centering
    \includegraphics[width=\linewidth]{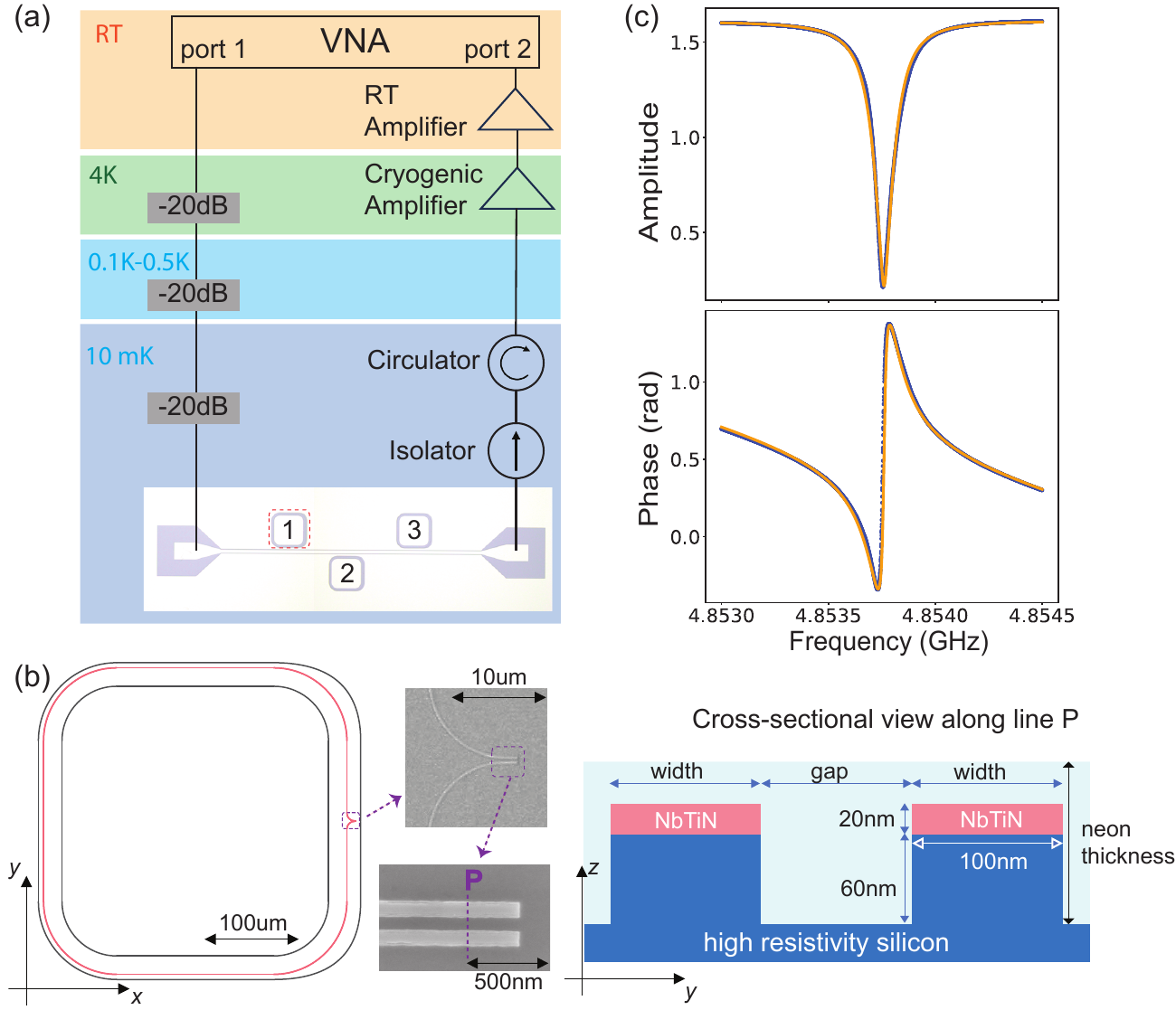}
    \caption{
        (a) Schematic overview of the electrical setup within the dilution refrigerator and at room temperature (RT). Input lines are attenuated with attenuators by $60\,\text{dB}$ in total, with output lines equipped with an LNF-LNC4\_8C amplifier at $4\,\text{K}$ and an LNF-LNR4\_8ART amplifier at RT. The diagram also includes an optical micrograph of the measured device on the $10\,\mathrm{mK}$ plate, showing three resonators—Resonator~1, Resonator~2, and Resonator~3—all sharing a single microwave (MW) feed line.
        (b) Conceptual illustration of Resonator 1. The nanowire resonator is shown in pink. Two black rounded squares represent grounded regions, with both the interior of the squares and the surrounding exterior connected to ground. SEM images of the enlarged areas highlight the two ends of the resonator. A schematic illustration of vertical cross-sections along line P is also presented.
        (c) Transmission response of Resonator~1 measured at $10\,\mathrm{mK}$ with a Vector Network Analyzer (VNA). The blue lines represent the measured $S_{21}$ in linear amplitude and phase before the deposition of solid neon or electrons. The orange lines are constructed from the best fit using the method described in Refs.~\onlinecite{Probst2015-gs, Probst2024resonator}.}
    \label{fig:sample}
\end{figure}

We fabricated three resonators, all sharing a single feedline (Fig.~\ref{fig:sample}). The two ends of each resonator meet with a small gap, where the electric field is maximized, enabling strong coupling between the electron's dipole moment and the electric field. Both the widths and the gaps were designed to be $100\,\mathrm{nm}$ for Resonator~1, $200\,\mathrm{nm}$ for Resonator~2, and $300\,\mathrm{nm}$ for Resonator~3. The resonance frequency of Resonator~1 was measured to be $f_\mathrm{r} = 4.81\,\mathrm{GHz}$ with $Q_\mathrm{ext} = 3.7 \times 10^4$, showing good agreement with simulations (Appendix~\ref{sec:TLC} and Table~\ref{reso_table}). $Q_\mathrm{int} = 2.3 \times 10^5$ and $Q_\mathrm{tot} = 3.2 \times 10^4$ were measured at $10\,\mathrm{mK}$. These values were obtained using an input power of $-21\, \text{dBm}$, corresponding to an estimated intracavity average photon number of $\langle n_{\text{ph}} \rangle \approx 10^5$. The resonance frequency of Resonator~2 was measured to be \(f_\mathrm{r} = 5.91\,\mathrm{GHz}\). Although both the width and gap were nominally designed to be \(200\,\mathrm{nm}\), over-etching appears to have reduced the resonator width, resulting in better agreement with simulation results for a width-gap configuration in the range of \(150\)–\(150\,\mathrm{nm}\) (Table~\ref{reso_table} and Appendix~\ref{sec:TLC}).  Other parameters for Resonators~2 and 3 are provided in Table~\ref{reso_table}. For all the resonators, the total loss rate $\kappa/2\pi= f_\mathrm{r} /Q_\mathrm{tot} \approx 0.1\,\mathrm{MHz}$.  
\begin{table*}[t]
\caption{\label{reso_table}COMSOL simulations and experimental data measured at $10\,\mathrm{mK}$ without neon or electrons. Intended width-gap dimensions for Resonators~1, 2, and 3 were 100\,nm/100\,nm, 200\,nm/200\,nm, and 300\,nm/300\,nm, respectively. For Resonator~2, simulation results for width-gap dimensions of 150\,nm/150\,nm, which are closer to those of the actual fabricated sample, are also included. See Appendix~\ref{sec:3D_RF_COMSOL} for more details of the simulation. For the experimental data, a power of $-21\,\text{dBm}$ generated from the VNA was used, corresponding to an estimated photon number in the resonator of $\braket{n}_\mathrm{ph} \approx 10^5$. For power dependence measurements, see Appendix~\ref{Power_Qint}.} 
\begin{ruledtabular}
\begin{tabular}{ccc@{\hspace{8pt}}|cc@{\hspace{6pt}}|cccc}

\multicolumn{3}{c@{\hspace{8pt}}|}{Resonators} & \multicolumn{2}{c@{\hspace{6pt}}|}{COMSOL Simulation} & \multicolumn{4}{c}{Experimental Data} \\ \hline
Resonator & Width/Gap (nm)& Length (\(\mu\text{m}\)) & \( f_\text{r} \) (GHz) & \( Q_{\text{ext}} \) & \( f_\text{r} \) (GHz) & \( Q_{\text{int}} \)  & \( Q_{\text{ext}} \)  & \( Q_{\text{tot}} \) \\

\hline
Resonator 1 & 100/100 & 1450 & 4.72 & \( 4.0 \times 10^4 \) & 4.81 & \( 2.3 \times 10^5 \) & \( 3.7 \times 10^4 \) & \( 3.2 \times 10^4 \) \\ 
Resonator 2 & 200/200 & 1450 & 6.42 & \( 2.7 \times 10^4 \) & 5.91 & \( 1.8 \times 10^5 \) & \( 2.5 \times 10^4 \) & \( 2.2 \times 10^4 \) \\ 
            & 150/150 & 1450 &  5.66 & \( 3.3 \times 10^4 \) &   &     &  &  \\ 
Resonator 3 & 300/300 & 1450 & 7.66 & \( 2.0 \times 10^4 \) & 6.98 & \( 1.7 \times 10^5 \) & \( 3.8 \times 10^5 \) & \( 1.2 \times 10^5 \) \\ 
\end{tabular}
\end{ruledtabular}
\end{table*}

We note that this device design intentionally omitted  DC bias lines to avoid degrading the $Q_\mathrm{int}$ of the NbTiN resonator, as the primary objective of this work was to establish a performance baseline of the resonators with solid neon and electrons.

\section{\label{sec:NeAndEDeposition}neon and electron deposition}

We deposited neon at $25\,\mathrm{K}$. Based on the resonance frequency shifts of Resonator~1 and Resonator~2, measured at $10\,\mathrm{mK}$ as $-0.94\%$ and $-0.86\%$, respectively, we estimated that the neon thicknesses in the regions of Resonator~1 and Resonator~2 are $160\,\mathrm{nm}$ and $270\,\mathrm{nm}$ at most, respectively (see Appendix~\ref{sec:neon_electron_sim}). From the volume of neon gas recovered after warming up, we estimated that the total amount of neon deposited in the experimental cell was $0.2\,\mathrm{mol}$. If we assume that the neon was deposited uniformly from the bottom of the experimental cell (surface area = $14\,\mathrm{cm}^2$), the expected neon thickness would be $1\,\mathrm{mm}$. This estimate is inconsistent with the resonance frequency shifts, suggesting that neon is primarily deposited on the walls of the experimental cell and capillary lines. This spread of neon to these surfaces may occur due to triple point wetting during the passage through the triple point when depositing solid neon~\cite{leiderer2025surface,migone1986triple}.  Aside from the thickness estimation, we also found that compared to the bare resonator,  \(Q_\mathrm{int}\) measured at $10\,\mathrm{mK}$ and at low photon number is increased with the presence of neon (Appendix~\ref{Power_Qint}). This must be attributed to the fact that the higher dielectric constant of neon reducing the participation ratio in the region where the TLS exists.

\begin{figure}[h!]
    \centering
    \includegraphics[width=1\linewidth]{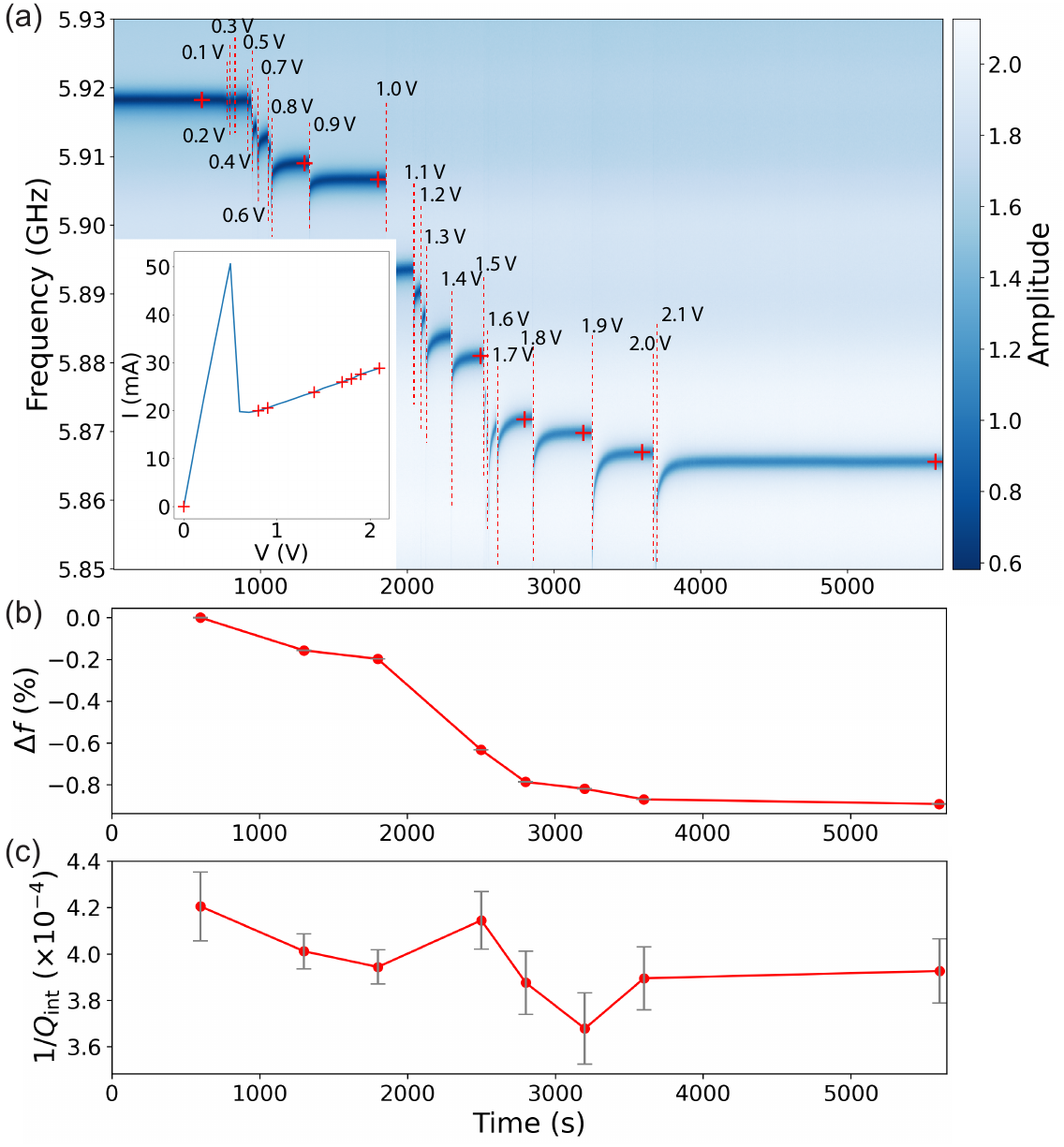}
    \caption{(a) Resonance peaks of Resonator~2 measured at $3.4\,\mathrm{K}$ over time. The vertical red dashed lines indicate the moments when electrons were emitted, with the applied filament voltages shown next to each. The red crosses mark the resonance peaks used to extract the resonance frequency shift and $Q_\mathrm{int}$ in (b) and (c), respectively. The inset displays the I-V curve of the filament. (b) The resonance frequency shift $\Delta f$, where the errors are smaller than the data points. (c) The inverse quality factor $1/Q_\mathrm{int}$. The error bars correspond to 95\% confidence. }
    \label{fig:e_deposition}
\end{figure}

After depositing neon, we proceeded with electron deposition by sending a voltage pulse to a filament placed in the vacuum cell. Figure~\ref{fig:e_deposition}(a) shows the resonance frequency shifts measured at $3.4\,\mathrm{K}$ by increasing the height of the voltage pulse (duration \(\approx 3\,\text{s}\)) applied to the filament. As shown in Fig.~\ref{fig:e_deposition}(b), the resonance frequency decreases with increasing electron number, while no change is observed in $Q_\mathrm{int}$, as shown in Fig.~\ref{fig:e_deposition}(c). The maximum measured frequency shift was $-0.9\%$.

The decrease in resonance frequency with increasing electron number on the neon surface is attributed to the surface conductivity of the two-dimensional electron layer. To illustrate this, we first consider the  Drude model:
\(
\sigma^{\mathrm{2D}} = \frac{e^2 n_e \tau}{m_e} \frac{1}{1 + i\omega \tau},
\) where \(e\) is the elementary charge, \(n_e\) is the surface electron density, \(\tau\) is the scattering time, \(m_e\) is the electron mass, \(\omega\) is the angular frequency of the microwave signal, and \(i\) is the imaginary unit. The highest reported \(\tau\) for electrons on solid neon is \(4.7\,\mathrm{ps}\), based on low-frequency mobility measurements at \(4.2\,\mathrm{K}\) \cite{Kajita1984-zr, Kajita1985-tz}. A shorter value of \(1.9\,\mathrm{ps}\) has also been reported for rougher surfaces \cite{Kajita1984-zr}. We simulated the resonance frequency shift \(\Delta f\) and the electron-induced loss \(1/Q_e\), incorporating surface conductivity based on the Drude model (Appendix~\ref{sec:simu_with_e}). The simulations show that a shift of \(-0.9\%\) corresponds to \(n_e = 0.4\,(0.8) \times 10^9\,\mathrm{cm}^{-2}\) and \(1/Q_e = 4.0\,(3.9) \times 10^{-3}\) for \(\tau = 4.7\,(1.9)\,\mathrm{ps}\). Since \(\tau\) is not directly measured and the Drude model does not accurately capture the measured \(Q_\mathrm{int}\), as discussed below, these values of \(n_e\) should be interpreted as indicative. The total internal quality factor is given by \(1/Q_\mathrm{int} = 1/Q_{\mathrm{int},0} + 1/Q_e\), where \(Q_{\mathrm{int},0}\) is the value without electrons. The simulated \(1/Q_\mathrm{int}\) is nearly an order of magnitude larger than the experimentally measured value, which remains around \(4 \times 10^{-4}\) at \(3.4\,\mathrm{K}\) and shows no dependence on electron density (Fig.~\ref{fig:e_deposition}(c)). This discrepancy suggests that the Drude model fails to capture the measured $Q_\mathrm{int}$. In Appendix~\ref{sec:simu_with_e_Lorentz}, we explore a model that incorporates surface disorder via harmonic confinement, which can localize electrons. The results show that in certain parameter regimes, this approach can account for the observed frequency shift without a significant reduction in $Q_\mathrm{e}$. These findings suggest that the limitations of the Drude model may arise from localization effects induced by surface roughness on a length scale smaller than the scattering length. In addition, the difference in the neon thickness observed between the two resonators can be attributed to large-scale nonuniformity of the neon deposition across the chip. 

After depositing electrons at \(3.4\,\mathrm{K}\), the sample was cooled down to \(10\,\mathrm{mK}\), where two-tone spectroscopy was performed (Appendix~\ref{2_tone}) to confirm that the electrons remained trapped during the cooling process.

\section{\label{sec:Magnet}magnetic field gradient}

\begin{figure}[H]
    \centering
    \includegraphics[width=\linewidth]{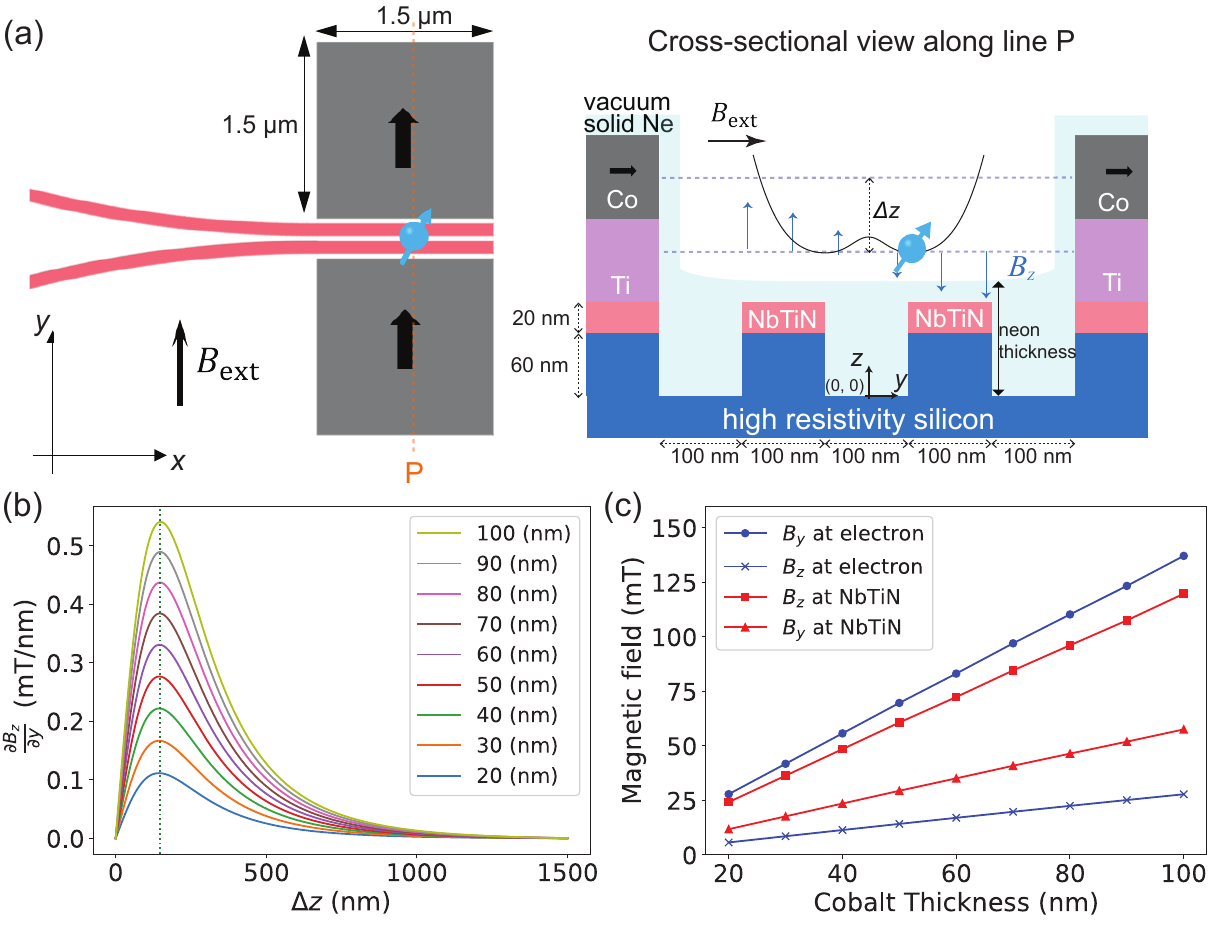}
    \caption{(a) Addition of Co magnets (gray) near the gap between the two ends of Resonator~1. In the cross-sectional view, the black line schematically represents the electrical potential experienced by the electron. (b) The magnetic field gradient in the \(z\)-direction along the \(y\)-axis, \(\frac{\partial B_z}{\partial y}\), is shown as a function of the vertical distance \(\Delta z\) between the center position of the Co magnet and the electron, for various Co thicknesses ranging from $20\,\text{nm}$ to $100\,\text{nm}$. (c) The absolute values of the offsets of the \(y\)- and $z$-components of the magnetic field at the electron’s position (i.e., at the potential minima), as well as the \(y\)- and \(z\)-components at the position of the NbTiN resonator, are plotted as functions of the Co thickness. For each Co thickness, the Ti thickness is optimized to maximize \(\frac{\partial B_z}{\partial y}\).}
    \label{fig:micro_magnet}
\end{figure}

The sample used in this work consists exclusively of resonators. As depicted in Fig.~\ref{fig:micro_magnet}, we propose to add two Co/Ti/NbTiN electrodes at both ends of Resonator~1, on its upper and lower sides. Cobalt (Co) serves as a ferromagnetic layer, titanium (Ti) as an adhesion layer, and NbTiN as part of the resonator. These electrodes are designed to serve three roles: generating local magnetic field gradients, tuning the electron's confinement potential through the application of DC voltage to the electrodes, and performing Electric Dipole Spin Resonance (EDSR)~\cite{Nowack2007,Tokura2006,Pioro-Ladriere2008,Dijkema2025} by applying microwave (MW) signals to one of the electrodes (Sec.~\ref{Qubit_gate}).

In the charge-qubit experiments reported in Refs.~\onlinecite{Zhou2022-nk,Zhou2023-iw}, a single electron trapped on solid neon shows a spectrum that qualitatively equals that seen in semiconductor double quantum dots (i.e., the energy of the charge state follows $\sqrt{\epsilon^2 + 4t_c^2}$, where $\epsilon$ is the charge detuning and $t_c$ is the effective tunnel coupling~\cite{Jennings2024-sb}), which may be due to the surface roughness of the solid neon~\cite{Kanai2024-bo}. Here, we assume that the two minima of the electrical potential are created at the edges of each end of the resonator. This assumption is plausible because these points are where the attraction towards the surface becomes maximum due to the image potential created by the resonators. This configuration is illustrated in Fig.~\ref{fig:micro_magnet}, and the distance between the two minima, i.e., the inter-dot distance $d = 100\,\text{nm}$.

By applying an external magnetic field $B_\mathrm{ext}$ along the $y$-axis, the Co parts are magnetized along the same axis and induce local magnetic fields. We present the numerical calculation results of the induced local magnetic fields and their gradients~\cite{Goldman} in Fig.~\ref{fig:micro_magnet}. Simulation details are given in Appendix~\ref{Mag_sim}. 

Figure~\ref{fig:micro_magnet}(b) shows the \( z \)-direction magnetic field gradient along the \( y \)-axis, \( \frac{\partial B_z}{\partial y} \), as a function of the distance between the \( z \)-center position of the Co magnet and the position of the electron, $\Delta z$. We found that the \( z \)-direction magnetic field gradient becomes maximum when $\Delta z \approx 146\,\text{nm}$ (vertical dashed line in Fig.~\ref{fig:micro_magnet}(b)). For example, with a Co thickness of $65\,\text{nm}$, a Ti thickness of $196\,\text{nm}$ ($126\,\text{nm}$), a neon thickness of $160\,\text{nm}$ ($90\,\text{nm}$), and an electron floating height of $2.5\,\text{nm}$, $\Delta z = 146\,\text{nm}$ can be achieved. The value of $\Delta z$ can be adjusted even after sample fabrication by modifying the neon thickness. With this configuration, we obtain \( \frac{\partial B_z}{\partial y} = 0.36\,\text{mT/nm} \), resulting in a magnetic field gradient in the \( z \)-direction between the two potential minima, expressed in radian frequency as \( b_\perp = \frac{g \mu_B}{\hbar} \frac{\partial B_z}{\partial y} \cdot d = 2\pi \cdot 1\,\mathrm{GHz} \), where \( g \) is the free electron \( g \)-factor, \( \mu_B \) is the Bohr magneton and \( \hbar \) is the reduced Planck constant. This magnetic field gradient strength is sufficient to reach the strong coupling regime for spin-photon coupling, as discussed in Sec.~\ref{sp_couping}, and to achieve high-fidelity qubit gates, as discussed in Sec.~\ref{Qubit_gate}.

Figure~\ref{fig:micro_magnet}(c) shows the magnetic field offsets along the \( z \)- and \( y \)-axes introduced by the Co magnets at the position of the NbTiN resonators, which are at most \( B_z = 60\,\text{mT} \) and \( B_y = 130\,\text{mT} \), respectively. These values are low enough not to degrade the quality factor of the resonators for a NbTiN thickness of $20\,\text{nm}$~\cite{Kroll2019-eo,Samkharadze2016-xh}. Small fabrication-induced displacements have little effect on this result, see Appendix~\ref{Mag_sim}. The \( B_y \) experienced by the electron at the positions of the potential minima is \( B_y = 90\,\text{mT}\) for the same configuration as discussed before (Co thickness of $65\,\text{nm}$ and $\Delta z \approx 146\,\text{nm}$). Thus, the total magnetic field that defines the Zeeman splitting in radian frequency is expressed as \( b_\parallel \approx \frac{g \mu_B}{\hbar} (B_y + B_\mathrm{ext}) \). Thus, to set the Zeeman splitting equal to the resonator resonance frequency of $4.8\,\mathrm{GHz}$, we need to apply \( B_\mathrm{ext} = 80\,\text{mT} \), which is sufficient to magnetize Co and perform qubit measurements~\cite{Samkharadze2018,Harvey-Collard2022-jh,Dijkema2025}. In the presence of a magnetic field gradient, charge noise—which causes fluctuations in the charge detuning $\epsilon$—could lead to fluctuations in the spin splitting. This makes the spin qubit sensitive to charge noise. Here, two Co magnets are placed symmetrically with a narrow gap relative to the electron positions. As a result, the Zeeman splitting at the two potential minima becomes equal, and the effect of charge noise at $\epsilon=0$ is largely suppressed.

\section{Spin-photon coupling and Loss Rate}
\label{sp_couping}

In this section, we estimate the spin-photon coupling strength, evaluate the loss rates of the spin state and the resonator, and discuss the feasibility of reaching the spin-photon strong coupling regime. For this purpose, we use the parameters of Resonator~1: an inductance of $L = 139\,\mathrm{nH}$, validated through experimental characterization (Appendix~\ref{K_in}) and the measured resonance frequency $f_\mathrm{r}=\omega_\mathrm{r}/2\pi=4.81$~GHz. The characteristic impedance of the resonator is then calculated as $Z_0=2f_\mathrm{r}L=1337~\Omega$.

According to Ref.~\onlinecite{Samkharadze2016-xh}, the zero-point fluctuation (ZPF) of the rms voltage amplitude between the two ends of the resonator can be calculated as:
\begin{equation}
    V_0 = \frac{1}{\sqrt{2}} \frac{2L}{\pi} \sqrt{ \frac{ 2 \hbar \omega_r}{L}} \omega_r = 12 \,\mu\mathrm{V},
\end{equation}
and thus the charge-photon coupling strength can be calculated as 

\begin{equation}
    \frac{g_c}{2\pi} = \frac{\alpha  e V_0 \cos \theta}{h}, \label{eq:gc}
\end{equation}
where $\theta=\arctan(\epsilon/2t_c)$ and $e V_0 /h = 3\,\mathrm{GHz}$. Here, we use \(\alpha = 0.05\). This yields \(g_c / 2\pi = 150\,\mathrm{MHz}\) at \(\epsilon = 0\), a value comparable to those measured in semiconductor quantum dots using NbTiN resonators with similar designs~\cite{Samkharadze2018,Dijkema2025}. Moreover, COMSOL simulations indicate that \(\alpha\) exceeds 0.1 for $d=100$~nm, when the neon layer thickness is below $300\,\text{nm}$, suggesting that the chosen value does not lead to an overestimation.

The high-frequency noise-induced loss rate is given by $\gamma =  \gamma_1 /2+\gamma_\phi$, where   $\gamma_1=2\pi/T_1$, $\gamma_\phi=2\pi/T_\phi$, \(T_1\) is the energy relaxation time, and \(T_\phi\) is the pure dephasing time. The quasistatic noise-induced loss rate is given by~\footnote{In analogy with Ref.~\citenum{Benito2019-pi}, the decay due to quasi-static noise is assumed to follow a Gaussian form, $\exp(-(\gamma^*t)^2)$.} \(
\gamma^* =\frac{1}{T_2^*}
\), where \(T_2^*\) is the inhomogeneous (quasistatic) dephasing time. We denote the loss rates for spin and charge by adding subscripts  ``\(\mathrm{s}\)'' and ``\(\mathrm{c}\)'', respectively, i.e., \(\gamma_{\mathrm{s}}\), \(\gamma_{\mathrm{s,1}}\), \(\gamma_{\mathrm{s,\phi}}\) and \(\gamma_{\mathrm{s}}^*\) for spin, and \(\gamma_{\mathrm{c}}\), \(\gamma_{\mathrm{c,1}}\), \(\gamma_{\mathrm{c,}\phi}\) and \(\gamma_{\mathrm{c}}^*\) for charge. The coherence times are denoted as \(T_{\mathrm{c},1}\), \(T_{\mathrm{c},\phi}\), \(T_{\mathrm{c},2}^*\), and \(T_{\mathrm{s},1}\), \(T_{\mathrm{s},\phi}\), \(T_{\mathrm{s},2}^*\), respectively.

To assess the impact of different loss mechanisms, we consider three representative scenarios, summarized in Table~\ref{table:loss_rate} and labeled as ``thermal,'' ``$^{\mathrm{nat}}$Ne,'' and ``$^{22}$Ne,'' corresponding to Scenarios~1, 2, and~3, respectively. Each scenario reflects a distinct set of assumptions regarding charge and spin loss rates. In \textbf{Scenario~1 (``thermal'')}, we employed charge loss rates of intermediate magnitude, as experimentally reported in Ref.~\onlinecite{Li2025-em}, representative of typical values found in previous studies. For the spin loss rate due to high-frequency noise, we used a value determined by thermal magnetic noise in copper. Although this effect is expected to be negligibly small---our electrodes are made of superconducting material, which exhibits minimal thermal magnetic noise, and both the printed circuit board and the experimental cell, though made of copper, are located far from the electrons---we include this case to evaluate a worst-case scenario. In \textbf{Scenarios~2 (``$^{\mathrm{nat}}$Ne'')} and \textbf{3 (``$^{22}$Ne'')}, we employed the smallest experimentally measured charge loss rates reported in Ref.~\onlinecite{Zhou2023-iw}. For the spin loss rates, we used values determined by hyperfine interactions with nuclear spins~\cite{Chen2022-on}. Scenario~2 corresponds to natural neon, whereas Scenario~3 assumes isotopically purified neon containing only $1\,\text{ppm}$ of $^{21}\mathrm{Ne}$.

\begin{table*}[htbp]
\centering
\caption{The table lists the charge and spin loss rates assumed in three different scenarios.
The charge loss rate values are experimentally measured at the charge sweet spot (\(\epsilon = 0\)) as reported in Refs.~\onlinecite{Zhou2023-iw,Li2025-em}. 
The spin loss rates are theoretical estimates from Ref.~\onlinecite{Chen2022-on}, where three scenarios are considered, 
each dominated by a different source of spin decoherence: thermal magnetic noise, hyperfine interaction with nuclear spins in natural neon 
(\(^{\mathrm{nat}}\text{Ne}\)), or hyperfine interaction in isotopically enriched \(^{22}\text{Ne}\) containing $1\,\text{ppm}$ of \(^{21}\text{Ne}\). For \(\gamma_\mathrm{s}\), as the relaxation time \(T_{\mathrm{s},1}\) is expected to be long~\cite{Chen2022-on} , $\gamma_\mathrm{s}\approx \gamma_{\mathrm{s},\phi}$. For \(\gamma_\mathrm{c}\), the measurements reported in Refs.~\onlinecite{Zhou2023-iw,Li2025-em} show that $\gamma_\mathrm{c} \approx \gamma_\mathrm{c,1}/2$.}
\begin{tabular}{|c|c|c|c|}
\hline
& Scenario 1 (thermal)& Scenario 2 ($^\mathrm{nat}$Ne)& Scenario 3 ($^\mathrm{22}$Ne)\\
\hline
\multirow{2}{*}{$\gamma_\mathrm{c}/2\pi$}& 
$\frac{1}{2T_{\mathrm{c},1}} =\frac{1}{2 \cdot 11.6\,\mu\mathrm{s}} =43.1\,\mathrm{kHz}$& 
\multicolumn{2}{c|}{$ \frac{1}{2 T_{\mathrm{c},1}} =\frac{1}{2 \cdot 48.2\,\mu\mathrm{s}} =10.4\,\mathrm{kHz}$} \\
& (Exp. value reported in Ref.~\onlinecite{Li2025-em}) & 
\multicolumn{2}{c|}{(Exp. value reported in Ref.~\onlinecite{Zhou2023-iw})} \\
\hline

\multirow{2}{*}{$\gamma_\mathrm{c}^*$}& 
$\frac{1}{ T_{\mathrm{c},2}^*} = \frac{1}{8.2\,\mu\mathrm{s}} =122\,\mathrm{kHz}$& 
\multicolumn{2}{c|}{$\frac{1}{T_{\mathrm{c},2}^*} = \frac{1}{42.8\,\mu\mathrm{s}} =23.4\,\mathrm{kHz}$} \\
& (Exp. value reported in Ref.~\onlinecite{Li2025-em})& 
\multicolumn{2}{c|}{ (Exp. value reported in Ref.~\onlinecite{Zhou2023-iw})} \\
\hline

\multirow{2}{*}{$\gamma_\mathrm{s}/2\pi$}& 
$\frac{1}{T_{\mathrm{s},\phi}}=\frac{1}{0.17\,\mathrm{ms}}=5.88\,\mathrm{kHz}$& 
$\frac{1}{T_{\mathrm{s},\phi}}= \frac{1}{30\,\mathrm{ms}}=33.3\,\mathrm{Hz}$& 
$\frac{1}{T_{\mathrm{s},\phi}}= \frac{1}{81\,\mathrm{s}}=0.012\,\mathrm{Hz}$\\
& (Calc. value for thermal noise ~\cite{Chen2022-on}) & 
(Calc. value for $^\mathrm{nat}$Ne ~\cite{Chen2022-on}) & 
(Calc. value for $^\mathrm{22}$Ne ~\cite{Chen2022-on}) \\
\hline

\multirow{2}{*}{$\gamma_\mathrm{s}^*$}& 
\multicolumn{2}{c|}{$\frac{1}{T_{\mathrm{s},2}^*}= \frac{1}{0.16~\mathrm{ms}}=6.25\,\mathrm{kHz}$} & 
$\frac{1}{T_{\mathrm{s},2}^*}= \frac{1}{0.43\,\mathrm{s}}=2.33\,\mathrm{Hz}$\\
& \multicolumn{2}{c|}{(Calc. value for $^\mathrm{nat}$Ne ~\cite{Chen2022-on})} & 
(Calc. value for $^\mathrm{22}$Ne ~\cite{Chen2022-on}) \\
\hline
\end{tabular} \label{table:loss_rate}
\end{table*}

We set \( \epsilon = 0 \) from here onwards for simplicity and to leverage the long coherence time at the charge sweet spot. To measure the strong coupling between the electron spin state and a microwave photon in the resonator, we tune the spin and photon energies into resonance by setting the Zeeman splitting equal to the resonator resonance frequency, that is, the detuning of the spin state from the resonator \(\Delta_\mathrm{s} = b_\parallel - \omega_r = 0\). The spin-photon coupling is given by \(g_\mathrm{s} = \Lambda  g_\mathrm{c}\), where \(
\Lambda = \sin {\bar{\phi}},
\) where $ {\bar{\phi}} = (\phi_+ + \phi_-)/2$ is the spin-charge mixing angle and $\phi_{\pm} = \arctan{[b_{\perp} /(2t_\mathrm{c} \pm b_{\parallel})]}$~\cite{Benito2019-pi}. Due to spin-charge coupling, the effective high-frequency noise-induced and quasi-static spin loss rate at the charge sweet spot become~\cite{Benito2019-pi}
\begin{equation}
  \gamma_\mathrm{s}' = \Lambda^2 \gamma_\mathrm{c,} + (1-\Lambda^2) \gamma_\mathrm{s},
\end{equation}
and
\begin{equation}
       {\gamma^*_\mathrm{s}}'= \sqrt{  \left( \Lambda^2\frac{\gamma_\mathrm{c}^*}{b_\parallel}\right)^2+ \left( \frac{(\cos \phi_+ + \cos \phi_-) }{2}\gamma_\mathrm{s}^* \right)^2 },
\end{equation}
respectively.

The condition for spectroscopically observing strong spin--photon coupling is \(g_s \gg \gamma_\mathrm{s}', {\gamma^*_\mathrm{s}}', \kappa'\). Here, we examine whether this condition is satisfied for Scenario 1. To evaluate the performance of the system under experimentally feasible conditions, we consider the following set of parameters as an example: \(b_\parallel / 2\pi = \omega_r / 2\pi = 4.8\,\mathrm{GHz}\), \(\epsilon = 0\), \(2t_c / 2\pi = 8.0\,\mathrm{GHz}\), and \(b_\perp / 2\pi = 1\,\mathrm{GHz}\). Under these conditions, we obtain \(\Lambda = 0.19\), resulting in a spin--photon coupling strength of \(g_s / 2\pi = 28.5\,\mathrm{MHz}\). The total resonator decay rate, including Purcell decay induced by the coupling between the charge state and the resonator, is given by \(\kappa' = \kappa + \frac{g_c^2 \gamma_{\mathrm{c}}}{\Delta_\mathrm{c}^2}\), where \(\Delta_\mathrm{c} = \sqrt{\epsilon^2 + 4t_\mathrm{c}^2} - \omega_r\) is the detuning of the charge state from the resonator~\cite{Benito2017-ok}. For the parameters considered here, \(\kappa'/2\pi \approx \kappa/2\pi = 0.1\,\mathrm{MHz}\), as determined experimentally (see Sec.~\ref{sec:NWresonator}). At the charge sweet spot, the effective quasi-static spin loss rate is not influenced by quasi-static charge noise. The spin loss rate due to high-frequency noise is calculated to be \(\gamma_\mathrm{s}' / 2\pi \approx 7\,\mathrm{kHz}\). These parameters satisfy the condition for strong coupling. The corresponding cooperativity is estimated to be \(C = g_\mathrm{s}^2 / (\gamma_\mathrm{s}' \kappa') \gtrsim 10^6\), confirming the realization of the strong spin--photon coupling regime.

\section{Qubit gates}
\label{Qubit_gate}

\begin{figure}[H]
    \centering
    \includegraphics[width=\linewidth]{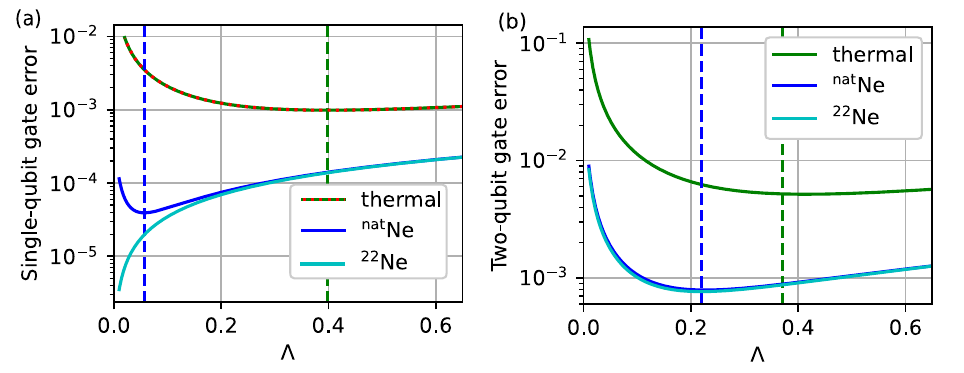}
    \caption{(a) Single-qubit gate error as a function of the spin--charge coupling strength \(\Lambda\). For Scenario 1, the green and red dotted lines show \(1 - F_1(0)\) and \(1 - \overline{F_1}\), with the minimum at \(\Lambda = 0.40\) (vertical green dashed line). For Scenarios 2 and 3, the blue and cyan lines show \(1 - F_1(0)\), respectively. For Scenario 2, the error is minimized at \(\Lambda = 0.057\), as indicated by the vertical blue dashed line. (b) iSWAP gate error \(1 - F_2\) as a function of \(\Lambda\), calculated from Eq.~\ref{eq:iSWAP}, for Scenario 1 (green line) with a minimum at \(\Lambda = 0.37\) (vertical green dashed line), and for Scenarios 2 and 3 (blue and cyan lines) with a minimum at \(\Lambda = 0.22\) (vertical blue dashed line). The parameter \(\beta = \Delta_s / g_s\) is set to 10, and \(\kappa / 2\pi = 0.1\,\mathrm{MHz}\) in all cases.}
    \label{fig:fidelity}
\end{figure}

In this section, we investigate the optimal experimental conditions for both single-qubit gates and an entangling two-qubit gate. 
To realize single-qubit gates, we can employ EDSR~\cite{Nowack2007,Tokura2006,Pioro-Ladriere2008,Dijkema2025} by applying MW signal to one of the Co/Ti/NbTiN electrodes and modulating the position of the electron along the $y$ axis. During EDSR, we detune the resonator from both the spin and charge states to suppress decoherence through the resonator (\(\Delta_\mathrm{c}, \Delta_\mathrm{s} \gg 0\)). The EDSR Rabi frequency is determined by $f^\mathrm{s}
_\mathrm{R}=\Lambda f^\mathrm{c}_\mathrm{R}$, where $f^\mathrm{c}_\mathrm{R}$ is the charge Rabi frequency. Here, we use \( f^\mathrm{c}_\mathrm{R} = 10\,\mathrm{MHz} \), as reported in Ref.~\onlinecite{Zhou2022-nk}. In Ref.~\onlinecite{Zhou2022-nk}, the microwave for charge state operation is applied through the resonator, whereas in our case, it will be supplied via one of the Co/Ti/NbTiN electrodes which are independent from the resonator. Although this difference could lead to a higher $f^\mathrm{c}_\mathrm{R}$, we adopt this value as a lower bound. Here, we evaluate the average gate fidelity of a $\pi$ gate as a representative example of single-qubit gate fidelity. The qubit gate operation can be performed faster than the timescale of quasi-static noise and can be corrected through feedback operations ~\cite{Kim2022-jq,Berritta2024-bb, Nakajima2021-tf, Nakajima2020-hr, Shulman2014-br, Park2025-cy}. Therefore, we first consider only high-frequency noise here. The average fidelity over all possible input states can be calculated as~\cite{Benito2019-pi}
\begin{equation}
   F_1(\delta) = \frac{1}{6} \left[3 + e^{-2t_g \gamma_{s}'} + 2 e^{-t_g \gamma_{s}'} \cos(t_g \delta)\right], 
\end{equation}
where \( t_g = 1 / (2 f_\mathrm{R}^s) \) is the gate duration, and \(\delta\) is the detuning between the spin resonance frequency and the frequency of the MW to the Co/Ti/NbTiN electrode. $F_1$ reaches its maximum when \(\Lambda^2\) matches \( \gamma_\mathrm{s}/(\gamma_\mathrm{c} - \gamma_\mathrm{s}) \), where the ratio \( f_{\mathrm{R}}^s / \gamma_\mathrm{s}'\) is maximized. When quasi-static noise is also taken into account, with $\delta$ following a Gaussian distribution with standard deviation $\sqrt{2} {\gamma^*_\mathrm{s}}'$, the average fidelity becomes
\begin{equation}
   \overline{F}_1 = \frac{1}{6} \left[3 + e^{-2t_g \gamma_{s}'} + 2 e^{-t_g \gamma_{s}'} e^{-\frac{(t_g {\gamma^*_\mathrm{s}}')^2}{2}}\right].
\end{equation}
Figure~\ref{fig:fidelity}(a) shows $1-F_1(0)$ for Scenarios 1, 2, and 3 and $1- \overline{F}_1$ for Scenario 1 as a function of $\Lambda$. For all the three scenarios, since the quasi-static noise is not significantly larger than the high-frequency noise, including it does not lead to a noticeable degradation in gate fidelity.

Next, we consider a two-qubit gate. Two-qubit operations can be performed on two electrons coupled to a resonator with the same geometry as the one used in this work~\cite{Li2025-em}. Moreover, by unfolding the resonator, we can implement remote two-qubit gates with electrons positioned at each end of the resonator, as demonstrated in Refs.~\onlinecite{Dijkema2025, Bottcher2022-dp,Borjans2020-do,Mi2018-co,Harvey-Collard2022-jh}. Among the various remote two-qubit gates~\cite{Burkard2020-gy}, we focus on the iSWAP gate, which has been both theoretically analyzed~\cite{Benito2019-al, Warren2019-mn} and experimentally realized for semiconductor spin qubits~\cite{Dijkema2025}. For this gate, spins are coupled dispersively to the resonator via virtual photons, with the frequencies of the two qubits significantly detuned from that of the resonator (\(\Delta_s = \beta g_s\), where \(\beta \gg 1\)). In the dispersive regime, we can largely suppress the effect of photon decay. The fidelity of the iSWAP gate is calculated as follows~\cite{Benito2019-al}:
\begin{equation}
    F_2=1-\frac{2\pi}{5g_s }\left( 2 \gamma_\mathrm{s}' \beta  +\frac{\kappa}{\beta }  \right) \label{eq:iSWAP}.
\end{equation}

Since the quasi-static noise is not substantially greater than the high-frequency noise, we assume it to be negligible during two-qubit gate operations and take into account only the high-frequency noise. \(F_2\) reaches its maximum when \(\Lambda^2\) matches \((\gamma_\mathrm{s} + \kappa / 2\beta^2)/(\gamma_\mathrm{c} - \gamma_\mathrm{s})\). Figure~\ref{fig:fidelity}(b) shows $1-F_2$ as a function of $\Lambda$ for Scenarios 1, 2, and 3. For Scenario 1, where \(\gamma_\mathrm{s}\) is on the same order as \(\gamma_\mathrm{c}\) but still smaller, and satisfies \(\gamma_\mathrm{s} \gg \kappa / 2\beta^2\), \(F_2\) reaches its maximum when \(\Lambda^2 = \gamma_\mathrm{s} / \gamma_\mathrm{c}\) (vertical green dashed line in Fig.~\ref{fig:fidelity}(b)). For Scenarios 2 and 3, where \(\gamma_\mathrm{s} \ll \gamma_\mathrm{c}, \kappa / 2\beta^2\), \(F_2\) reaches its maximum when \(\Lambda^2 = \kappa / (\beta^2 2\gamma_\mathrm{c})\) (vertical blue dashed line in Fig.~\ref{fig:fidelity}(b)). At this point, \(F_2 = 1 - \frac{4\pi}{5} \sqrt{\frac{2\gamma_\mathrm{c} \kappa}{g_c^2}}\), which is independent of both the spin--photon coupling strength and the spin decoherence rate, and reduces to the same form as that derived for semiconductor spin qubits in Eq.~(7) of Ref.~\onlinecite{Benito2019-al}.
To evaluate the performance of the system under experimentally feasible conditions in the dispersive regime, we consider the following set of parameters as an example: \(\omega_r / 2\pi = 4.8\,\mathrm{GHz}\), \(\epsilon = 0\), \(2t_c / 2\pi = 8.0\,\mathrm{GHz}\), \(b_\perp / 2\pi = 1\,\mathrm{GHz}\), \(b_\parallel / 2\pi = 5.1\,\mathrm{GHz}\). Under these conditions, we obtain \(\Lambda = 0.2\) and \(\beta = 10\). As shown in Fig.~\ref{fig:fidelity}(b), at \(\Lambda = 0.2\), the single-qubit gate fidelity reaches 99.877\% for Scenario 1, 99.992\% for Scenario 2, and 99.996\% for Scenario 3, while the iSWAP gate fidelity reaches 99.34\% for Scenario 1 and 99.92\% for Scenarios 2 and 3.

Gate fidelities might be further improved by dynamically tuning \(2t_c\) (or $\epsilon$) via voltage gate pulses to optimize \(\Lambda\) independently for single-qubit and two-qubit gate operations. For instance, in Scenario 3 with isotopically purified neon, reducing \(\Lambda\) can further enhance the single-qubit gate fidelity to exceed 99.999\%. This can be achieved by increasing \(2t_c\); however, tuning to high \(2t_c\) may be challenging if it is determined by surface roughness, as discussed in Sec.~\ref{sec:NeAndEDeposition}. Alternatively, reducing \(b_\perp\) through device design also lowers \(\Lambda\), but this comes at the cost of a reduced spin Rabi frequency \(f_\mathrm{R}^\mathrm{s}\), which increases the gate time \(t_g\) and ultimately degrades fidelity. Moreover, at this level of high gate fidelity, instrumental factors such as phase noise in the control electronics become increasingly important. To fully benefit from isotopic purification, it is therefore crucial to also consider these instrumental limitations~\cite{Ball2016-zn}. Since the iSWAP gate fidelity is limited not by the spin loss rate but by the charge and resonator loss rates, isotopic purification of neon does not yield a significant improvement in the two-qubit gate performance, at least for the scheme considered here.

\section{Conclusion}

In conclusion, we experimentally demonstrated neon and electron depositions using NbTiN nanowire resonators, aiming for the realization of spin-photon coupling and high-fidelity spin qubit gates in the future. The presence of neon and electrons was confirmed by a decrease in the resonance frequency without compromising the resonator’s quality factor, supporting the suitability of NbTiN resonators for future qubit measurements. A closer examination of the resonator's response to electron deposition suggests that the Drude model fails to capture the system's behavior, likely due to electron localization induced by surface roughness in the neon. At present, the control over thickness and surface roughness in the neon deposition process is not well established and requires improvement. Additionally, we investigated the configuration of cobalt magnets needed for realizing spin-photon coupling and spin qubit gates. By combining the enhanced charge-photon coupling enabled by the high impedance of the NbTiN resonator with the calculated magnetic field gradient produced by the cobalt magnets, we estimated that spin-photon coupling can reach the strong coupling regime, allowing for the implementation of high-fidelity spin qubit gates.

Future work will involve the experimental implementation of integrated micromagnets and additional DC electrodes, as theoretically proposed in this work. To avoid degradation of the resonator’s quality factor in such implementations, mitigation strategies such as perforated ground planes~\cite{Kroll2019-eo} and DC filters~\cite{Mi2017-go,Harvey-Collard2020-dt} will be necessary. In this context, the high quality factor demonstrated here provides an important target and benchmark for evaluating the effectiveness of these approaches.

\begin{acknowledgments}
We acknowledge Yasunobu Nakamura and the Superconducting Quantum Electronics Research Team at RIKEN for their support in fabricating the sample. We thank Pasquale Scarlino, Maja Cassidy, and Kun Zuo for useful discussions. This work was supported by RIKEN-Hakubi program, RIKEN Center for Quantum Computing, JST-FOREST and Yazaki foundation. D. J. acknowledges support from the Air Force Office of Scientific Research (AFOSR) under Award No. FA9550-23-1-0636.
\end{acknowledgments}
\appendix


\section{SAMPLE FABRICATION}

Samples are fabricated on high-resistivity (\(\rho \geq 10\,\mathrm{k}\Omega\cdot\mathrm{cm}\)) (100)-oriented intrinsic silicon substrates. The substrates are cleaned in a piranha solution for 15 minutes and then rinsed with deionized (DI) water. To remove the surface oxide, they are dipped in buffered hydrofluoric acid (BHF) for 5 minutes, followed by thorough rinsing with DI water. A 20-nm-thick NbTiN thin film is prepared by reactive magnetron sputtering using an 8-inch alloy target with a weight ratio of \(\mathrm{Nb}:\mathrm{Ti} = 5:1\). During deposition, the substrate is kept at ambient temperature, with a DC sputtering current of $3\,\text{A}$ and a gas pressure of $2\,\text{mTorr}$ \(\mathrm{(Ar:N_2 = 100{:}37)}\). After cutting the wafer with the NbTiN thin film into $10 \times 10 \, \mathrm{mm}$ pieces, it undergoes oxygen plasma cleaning for 1 minute to remove organic contamination, followed by a 1 minute buffered HF (BHF) clean to eliminate any remaining residues. A 100-nm-thick layer of ZEP520A resist, diluted in anisole, is spin coated onto the substrate. The coated substrate is then baked at 180$^\circ$C for 3\,minutes. Electron-beam lithography (EBL) is carried out using an ELS-F125 system with an electron dose of 230\,$\mu$C/cm\textsuperscript{2} and a beam current of $1\,\text{nA}$ to define the microwave circuitry. After exposure, the resist is developed in ZED-N50 for 2 minutes, followed by a 30-second rinse in IPA. The pattern is transferred to the film by reactive ion etching (RIE) using a gas mixture of O\textsubscript{2} for 25 seconds and CF\textsubscript{4} at $50\,\text{sccm}$, $100\,\text{W}$, and $10\,\text{Pa}$ for\,75 seconds. This process results in the successful fabrication of the NbTiN resonator on the chip, with the underlying 60-nm Si layer also being etched.

\section{Simulation for Neon and Electron Deposition}
\label{sec:neon_electron_sim}

\begin{figure}
    \centering
    \includegraphics[width=\linewidth]{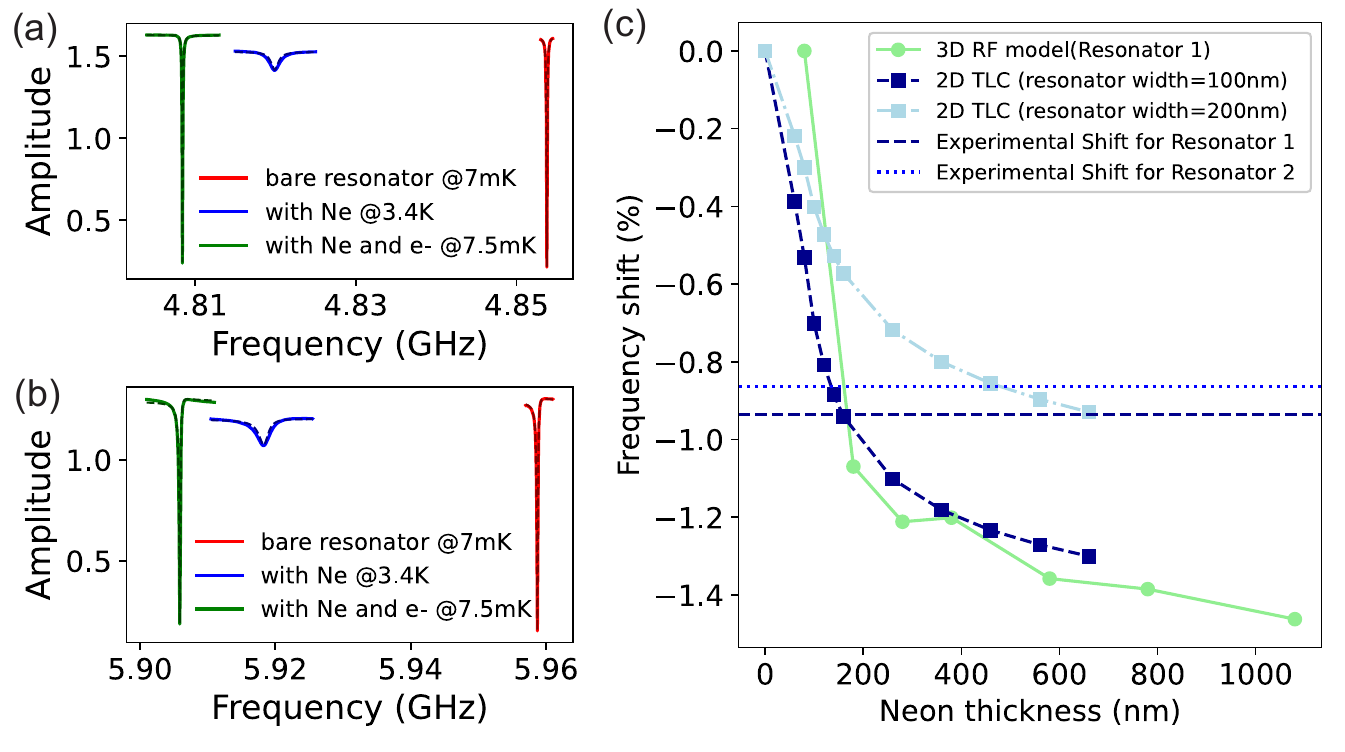}
    \caption{
        (a) Resonance peaks measured for the bare Resonator~1 at $7\,\mathrm{mK}$ (red), with neon at $3.4\,\mathrm{K}$ (blue), and with neon and electrons at $7.5\,\mathrm{mK}$ (green). The corresponding internal quality factors ($Q_\mathrm{int}$: $2.34 \times 10^5$, $4.66 \times 10^3$, $2.22 \times 10^5$), external quality factors ($Q_\mathrm{ext}$: $3.19 \times 10^4$, $4.33 \times 10^3$, $3.36 \times 10^4$), and resonance frequencies ($f_\mathrm{r}$: $4.854\,\mathrm{GHz}$, $4.82\,\mathrm{GHz}$, $4.808\,\mathrm{GHz}$) are given for red, blue, and green, respectively. (b) Resonance peaks measured for the bare Resonator~2 at $7\,\mathrm{mK}$ (red), with neon at $3.4\,\mathrm{K}$ (blue), and with neon and electrons at $7.5\,\mathrm{mK}$ (green). The corresponding internal quality factors ($Q_\mathrm{int}$: $1.91 \times 10^5$, $3.9 \times 10^3$, $1.42 \times 10^5$), external quality factors ($Q_\mathrm{ext}$: $2.16 \times 10^4$, $3.49 \times 10^3$, $1.97 \times 10^4$), and resonance frequencies ($f_\mathrm{r}$: $5.959\,\mathrm{GHz}$, $5.918\,\mathrm{GHz}$, $5.906\,\mathrm{GHz}$) are given for red, blue, and green, respectively. (a, b) The dotted lines represent the fits. (c) Resonance frequency shift obtained from the COMSOL simulation as a function of the neon thickness. The vertical dotted and dashed lines represent the maximum measured resonance frequency shifts of Resonators~1 and 2 caused by the neon and electron deposition, $-0.94\%$ and $-0.86\%$, respectively. Resonance frequency shifts for Resonator~1 are simulated using 3D RF model. }    
        
    \label{fig:neon_deposition}
\end{figure}

Figure~\ref{fig:neon_deposition}(a,b) represents resonant peaks measured for Resonators~1 and 2 with and without neon and electrons at $3.1\,\text{K}$ and around $7\,\text{mK}$, and Fig.~\ref{fig:neon_deposition}(c) shows the resonance frequency shifts calculated as a function of neon thickness using COMSOL simulations. The neon thickness is defined from the bottom of the etched Si substrate. For the “3D RF model” described in Appendix~\ref{sec:3D_RF_COMSOL}, since the resonator is defined on a 2D plane located 80\,nm above the bottom of the etched Si substrate,
 the graph starts from a neon thickness of $80\,\mathrm{nm}$.

The resonance frequency shift measured around $7\,\mathrm{mK}$ can be attributed to both the effects of neon and electron deposition. By comparing the experimental and simulation results, we can thus set an upper bound for the neon thickness. 

To obtain a reliable estimate of the thickness, we employed and compared two independent simulation methods: a 3D electromagnetic simulation using COMSOL (Appendix~\ref{sec:3D_RF_COMSOL}) and a simplified 2D transmission line calculator (TLC, Appendix~\ref{sec:TLC}). The 3D model captures detailed geometric and field distribution effects, while the TLC approach provides fast, intuitive access to the dependence on dielectric loading. Consistent results from both methods increase our confidence in the estimated neon thickness.

Thus, the neon thickness deposited on Resonator~1 was estimated to be $160\,\text{nm}$ in both models. For Resonator~2, the resonance frequency is closer to the simulation result for a width of $150\,\text{nm}$ than to that for $200\,\text{nm}$ (Table~\ref{reso_table}). Therefore, using the simulation result from the TLC (Appendix~\ref{sec:TLC}) for a width of $150\,\text{nm}$, the neon thickness was estimated to be $270\,\text{nm}$.

\subsection{ 3D RF model \label{sec:3D_RF_COMSOL}}

The simulation was performed using COMSOL Multiphysics v.~6.2. We began the simulation by creating a 3D model with the “Radio Frequency” module loaded for the physics setup, within which the “Electromagnetic Waves, Frequency Domain (emw)” interface is used. The study type was set to Frequency Domain and MUMPS direct solver was used to compute the response of the system subjected to harmonic excitation over a range of frequencies.

We use a 3D model with dimensions of \( 2000\,\mathrm{\mu m} \times 2000\,\mathrm{\mu m} \times 800\,\mathrm{\mu m} \), where the center of the resonator is roughly positioned at the origin. For simulating the resonator response in the absence of neon, the region \( 301.3 < z < 800\,\mathrm{\mu m} \) is defined as vacuum, and the region \( 0 < z < 300\,\mathrm{\mu m} \) is defined as silicon (the relative permittivity of silicon is set to \( 11.4 \)). Thirteen layers with thickness of $100\,\text{nm}$ were added from \( 300 < z < 301.3\,\mathrm{\mu m} \) each of which could be set as neon ($\epsilon_r=1.244$) or vacuum to simulate a neon layer of various thicknesses.

We set a 2D plane at \( z=300\,\mathrm{\mu m} \), where the resonators, the feedline, and the grounds are defined. The transmission of the electromagnetic field is solved using lumped ports 1 and 2, which are defined between each end of the feedline and the ground plane, each with a characteristic impedance of 50\,\(\Omega\). The transmission parameter \( S_{21} \) is then plotted as a function of the frequency \( f \) of the microwave signal sent through the feedline.

We account for the kinetic inductance of NbTiN by modeling it as a complex conductivity within the “Transition Boundary Condition” applied to the resonator. According to Ref.~\onlinecite{gao2008physics}, the current density \( \vec{J} \) can be expressed as
\[
\vec{J} =  \sigma \vec{E},
\]
where \( \vec{E} \) is the electric field, and the conductivity \( \sigma \) for a thin film is given by
\begin{equation}
\sigma = \frac{1}{i \mu_0 \omega \lambda^2}, \label{eq:sigma}
\end{equation}
where \( \omega = 2\pi f \), \( \mu_0 \) is the magnetic permeability, and \( \lambda \) is the magnetic penetration depth. In the COMSOL simulation, the electrical conductivity for the “Transition Boundary Condition” of the resonator is set according to Eq.~\ref{eq:sigma}. When the sample thickness is comparable to or smaller than the skin depth, as is the case here, we deselect the “Electrically thick layer” option and specify a layer thickness of \( D = 20\,\mathrm{nm} \).

The feedline and grounds are modeled as “Perfect Electric Conductor,” even though they are also made of NbTiN, because their kinetic inductance is expected to have a negligible impact on the resonator response for this geometry. The quality factor of the resonator computed in this COMSOL simulation is determined solely by the external quality factor, as we treat silicon as a lossless dielectric in the simulation.

Table~\ref{reso_table} shows the comparison between the measured data and the COMSOL simulation with \( \lambda = 390\,\mathrm{nm} \). Compared to Resonator~1, Resonators~2 and~3 exhibit deviations from the simulation results based on the intended width and gap dimensions. This is believed to be due to the width of Resonators~2 and 3 being smaller than intended, resulting in lower frequencies compared to the simulation.

\subsection{ ``Transmission Line Calculator''  COMSOL model \label{sec:TLC}}

We modeled the nanowire resonator as a coplanar waveguide using the ``Transmission Line Calculator'' COMSOL model. Although the actual resonator is not a straight nanowire, in this model, it is approximated as a transmission line aligned with the $x'$-axis (designating this extending direction as the $x'$-axis). Electromagnetic fields, specifically transverse electromagnetic (TEM) waves, propagate along the $x'$-axis. Here, we also assume that $x'$ components of electric and magnetic fields are small and the propagating mode is deduced
from separate magnetic and electric analyses.

The width of the resonator is defined along the $y'$-axis as 150\,nm and the thickness along the $z$ axis as 20\,nm, and it extends indefinitely along the $x'$-axis. In this model, we define metal, silicon, neon, and electrons in the $y'$-\(z\) plane. 

\subsubsection{Simulation of neon thickness  \label{sec:simu_Ne_TLC}}

To estimate the neon thickness, the neon thickness is considered in the \(z\)-direction, growing sequentially. We obtained the shunt capacitance per unit length, denoted as \( C_\mathrm{l} \). The frequency shift due to the presence of neon was then calculated using the formula:
\(
\Delta f = \sqrt{\frac{C_\mathrm{l}^\mathrm{w/o \, Ne}}{C_\mathrm{l}^\mathrm{w/ \, Ne}}} - 1
\), where \( C_\mathrm{l}^\mathrm{w/o \, Ne} \) and \( C_\mathrm{l}^\mathrm{w/ \, Ne} \) represent the shunt capacitance per unit length without and with neon, respectively.

\subsubsection{Simulation with the presence of electrons: Drude model  \label{sec:simu_with_e}}

To incorporate the effect of electrons, we first use the Drude model~\cite{Andrei1997Two-DimensionalSubstrates} based on the same 2D framework described above. The neon thickness is \(270\,\mathrm{nm}\), and the surface electrons are assumed to be located \(2.5\,\mathrm{nm}\) above the neon surface, spreading along the \(x'\)-\(y'\) plane. The equation of motion for an electron is given by
\begin{equation}
    m_e \left(\frac{\partial \mathbf{v}}{\partial t} + \frac{\mathbf{v}}{\tau} \right) = -e \mathbf{E}, 
    \label{eq:eq_motion_0}
\end{equation}
where \(\mathbf{E} = \mathbf{E}_0 \exp(i \omega t)\) is the electric field acting on the electron~\footnote{Note that we use the convention \(\exp(i \omega t)\) for the time dependence of fields. While the physics community commonly adopts \(\exp(-i \omega t)\), the engineering convention \(\exp(i \omega t)\) is used here for consistency with COMSOL.}, and \(\mathbf{v}\) is the electron velocity. Here, \(\tau\) is the scattering time, and \(m_e\) is the electron mass. The solution to Eq.~\ref{eq:eq_motion_0} is
\begin{equation}
    \mathbf{v} = -\frac{e \tau}{m_e} \frac{1}{1 + i \omega \tau} \mathbf{E}_0 \exp(i \omega t).
\end{equation}
The corresponding 2D current density is \(\mathbf{j} = -e n_e \mathbf{v} = \sigma^{\mathrm{2D}} \mathbf{E}_0 \exp(i \omega t)\), leading to the well-known Drude model:
\begin{equation}
    \sigma^{\mathrm{2D}} = \frac{e^2 n_e \tau}{m_e} \frac{1}{1 + i \omega \tau}.
    \label{eq:Drude_conductivity}
\end{equation}
To integrate this into the COMSOL model, the electron conductivity is defined as
\begin{equation}
     \sigma_{ij} = \frac{1}{\delta z} \frac{e^2 n_e \tau}{m_e}  
            \frac{1}{1 + i \omega \tau}
\end{equation}
for \(i = j = x'\) or \(i = j = y'\), and \(\sigma_{ij} = 0\) otherwise, where $\delta z = 1\,\text{nm}$ is the arbitrarily defined thickness of the area considered as the electron layer. To check the validity of this choice, simulations were also conducted with $\delta z = 2.5\,\text{nm}$, and it was confirmed that the differences in the results were negligibly small. Please note that, as the $x'$ component of the electric and magnetic fields is small, $\sigma_{y'y'}$ mainly contributes to the resonance peak change. We obtained the shunt capacitance per unit length and calculated the frequency shift as 
    \(
    \Delta f = \sqrt{\frac{C_\mathrm{l}^\mathrm{w/o \, e^-}}{C_\mathrm{l}^\mathrm{w/ \, e^-}}} - 1
    \), where \( C_\mathrm{l}^\mathrm{w/o \, e^-} \) and \( C_\mathrm{l}^\mathrm{w/ \, e^-} \) represent the shunt capacitance per unit length without and with electrons, respectively. The attenuation constant \(\alpha\) was extracted from the complex propagation constant \(\gamma = \alpha + i \beta\), computed in the simulation. The internal quality factor due to the electrons is then obtained via~\cite{Pozar1998}
\begin{equation}
    Q_e=\frac{\pi}{2 \alpha l}.
\end{equation}

\begin{figure}[h!]
    \centering
    \includegraphics[width=1\linewidth]{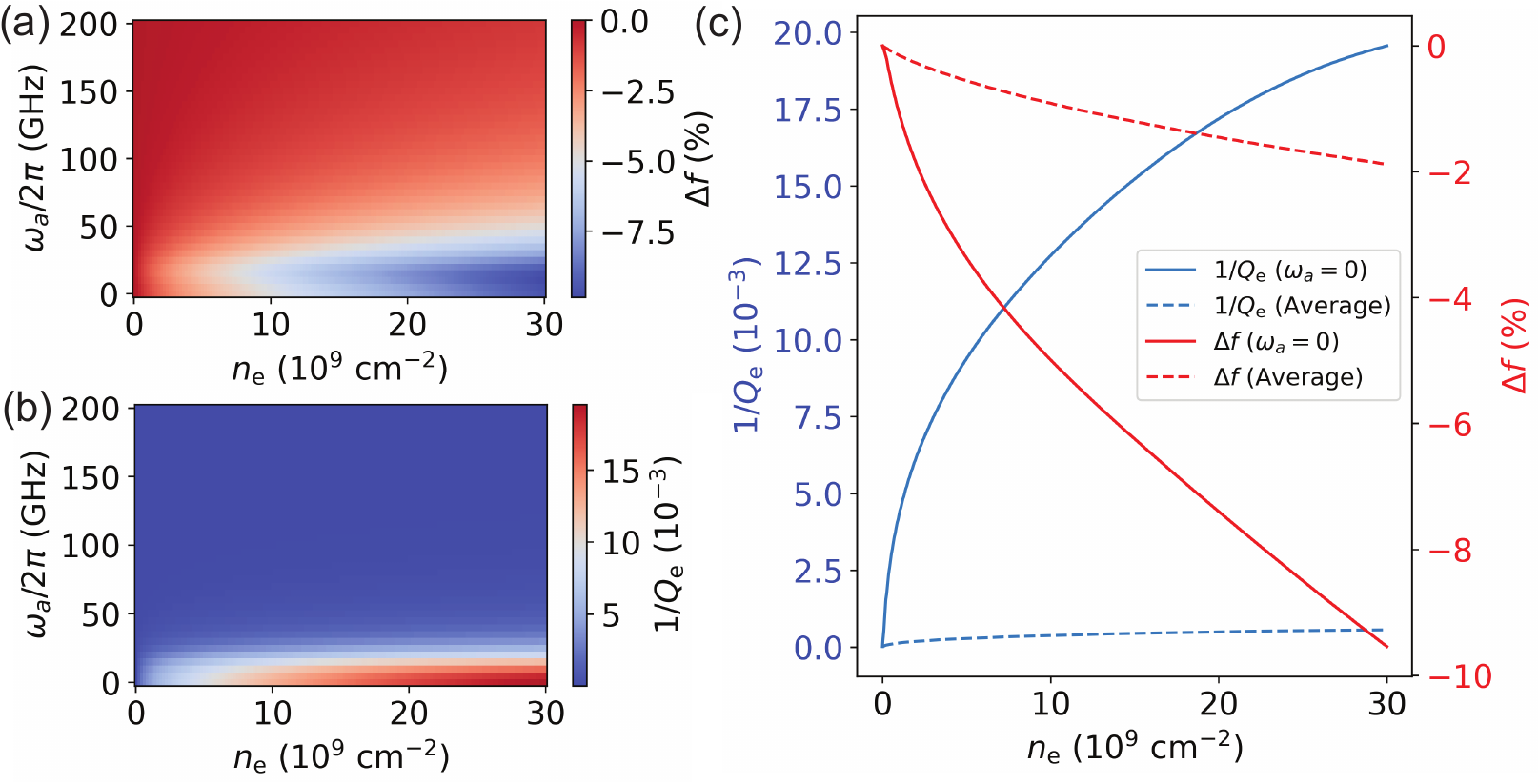}
    \caption{(a,b) The resonance frequency shift \(\Delta f\) in (a) and the inverse quality factor \(1/Q_e\) in (b), due to the presence of electrons, are obtained from simulations using Eq.~\ref{eq:cond_COMSOL_omega_a} as a function of the electron density \(n_e\) and the confinement frequency \(\omega_a/2\pi\). 
(c) Simulated \(1/Q_e\) (blue line) and \(\Delta f\) (red line) as functions of \(n_e\) for \(\omega_a/2\pi = 0\,\mathrm{GHz}\) (solid line) and average over \(\omega_a/2\pi \in [0, 200]\,\mathrm{GHz}\) with weights \(\propto\exp(
\hbar\omega_a/k_BT)\) (dashed line). 
}
    \label{fig:f_Q_Sim}
\end{figure}

Using this approach, we simulated the frequency shift \(\Delta f\) and electron-induced internal loss \(1/Q_e\) with \(\tau = 1.9\,\mathrm{ps}\), corresponding to the \(\omega_a = 0\) case in Fig.~\ref{fig:f_Q_Sim}. As discussed in the main text, the resulting \(1/Q_e\) from the simulation is nearly an order of magnitude larger than the experimentally measured value, which remains around \(4 \times 10^{-4}\) at \(3.4\,\mathrm{K}\) and shows no discernible dependence on electron density (Fig.~\ref{fig:e_deposition}(c)).

\subsubsection{Simulation with the presence of electrons: Lorentz model  \label{sec:simu_with_e_Lorentz}}

To explain the discrepancy between the measured and simulated internal quality factors, we consider electron localization induced by surface roughness. As one possible modeling approach, we assume that all electrons are confined in harmonic potential wells characterized by a frequency \(\omega_a\). The electron dynamics are then governed by the equation of motion:
\begin{equation}
    m_e \left(\frac{\partial \mathbf{v}}{\partial t} + \frac{\mathbf{v}}{\tau} \right) = -e \mathbf{E} - m_e \omega_a^2  \mathbf{x},
    \label{eq:eq_motion}
\end{equation}
where \(\mathbf{x}\) is the electron's position, taking the origin as the center of the harmonic potential it experiences, with a confinement frequency \(\omega_a\). Here, \(\tau\) is the scattering time, and \(m_e\) is the electron mass. The solution to Eq.~\ref{eq:eq_motion} is
    \begin{equation}
     \mathbf{v} =i \omega \mathbf{x}= -\frac{e \tau}{m_e} \frac{1}{1 + i \left( \omega - \frac{\omega_a^2}{\omega} \right) \tau} \mathbf{E}_0 \exp(i \omega t).
    \end{equation}
The resulting modified surface conductivity is given by:

\begin{equation}
    \sigma ^{\mathrm{2D}}= \frac{e^2 n_e \tau}{m_e} \frac{1}{1 + i \left( \omega - \frac{\omega_a^2}{\omega} \right) \tau}.
    \label{eq:2D_conductivity}
\end{equation}
To integrate this into the COMSOL model, the electron conductivity is now redefined as
\begin{equation}
     \sigma_{ij} = \frac{1}{\delta z} \frac{e^2 n_e \tau}{m_e}  
            \frac{1}{1 + i \left( \omega - \frac{\omega_a^2}{\omega} \right) \tau}. \label{eq:cond_COMSOL_omega_a}
\end{equation}
Using the same approach as described above, we simulated the frequency shift \(\Delta f\) and electron-induced internal loss \(1/Q_\mathrm{e}\) with \(\tau = 1.9\,\mathrm{ps}\), for trap frequencies \(\omega_a/2\pi\) ranging from \(0\) to \(200\,\mathrm{GHz}\) (Fig.~\ref{fig:f_Q_Sim}(a,b)). Introducing a finite \(\omega_a\) enhances the simulated resonance frequency shift around \(\omega_a \sim \sqrt{\omega/\tau}\), where the imaginary part of \(\sigma^{\mathrm{2D}}\) peaks. Meanwhile, the degradation of the simulated \(Q_{\mathrm{int}}\) is reduced as \(\omega_a\) increases, since the real part of \(\sigma^{\mathrm{2D}}\) decreases monotonically with increasing \(\omega_a\).

\begin{figure}[H]
    \centering
    \includegraphics[width=\linewidth]{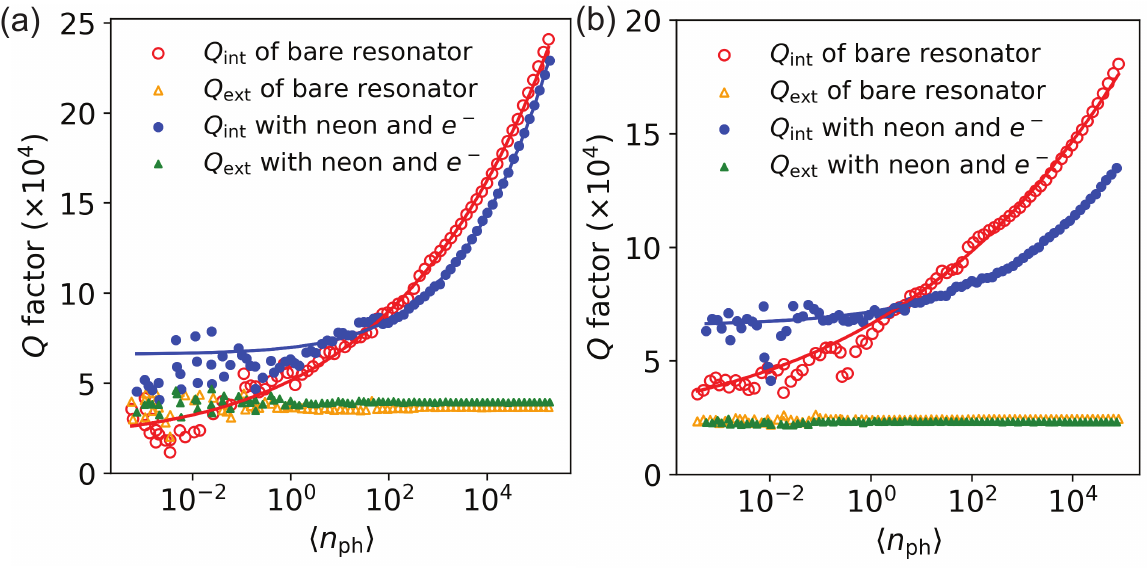}
   \caption{
Measured quality factor power dependence for Resonator~1 in (a) and Resonator~2 in (b) as a function of the number of photons in the resonators, \( \langle n_{\mathrm{ph}} \rangle \). The red open circles and blue-filled circles represent \( Q_\mathrm{int} \) of the bare resonator and the resonator with electrons and neon, respectively. The orange open triangles and green-filled triangles represent \( Q_\mathrm{ext} \) of the bare resonator and the resonator with electrons and neon, and we found that they are independent of \( \langle n_{\mathrm{ph}} \rangle \) with no observable changes before and after the deposition of electrons and neon. The red and blue solid lines are fits to \( Q_\mathrm{int} \) using Eq.~\ref{eq:TLS_loss}. (a) For Resonator~1, \( Q_{\mathrm{TLS, 0}}/F \) increases from \( (1.94\pm0.87)\times10^4 \) to \( (6.64\pm0.92)\times10^4 \), \( n_\mathrm{sat} \) increases from \( (1\pm5)\times10^{-3} \) to \( (3.0\pm1.2)\times10^{2} \), and \( \beta \) increases from \( 0.265\pm0.051 \) to \( 0.377\pm0.063 \) after the deposition of neon and electrons. Estimating \( Q_{\mathrm{other}} \) from the fit is difficult due to the lack of data at high photon numbers. However, we find that \( Q_{\mathrm{other}} \) changes from \( (2 \pm 49) \times 10^7 \) to \( (2 \pm 37) \times 10^7 \) after the deposition of neon and electrons. (b) For Resonator~2, \( Q_{\mathrm{TLS, 0}}/F \) increases from \( (2.2\pm2.2)\times10^4 \) to \( (6.8\pm1.6)\times10^4 \), \( n_\mathrm{sat} \) increases from \( (1\pm14)\times10^{-5} \) to \( (4.7\pm2.8)\times10^{2} \), and \( \beta \) increases from \( 0.190\pm0.060 \) to \( 0.256\pm0.059 \) after the deposition of neon and electrons. Additionally, \( Q_{\mathrm{other}} \) changes from \( (2\pm10)\times10^6 \) to \( (2\pm6)\times10^6 \) after the deposition of neon and electrons. For both resonators, the errors are estimated at the 95\% confidence level. The large uncertainties in \( n_{\mathrm{sat}} \) and \( Q_{\mathrm{other}} \) arise because we did not reach a high enough power to fully saturate the TLS. However, the values of \( n_{\mathrm{sat}} \) and \( Q_{\mathrm{other}} \) did not affect the fitting results of the other parameters.}
    \label{fig:Qint_vs_nph}
\end{figure}

To explain the experimentally observed absence of degradation in \( Q_\mathrm{int} \), we take the following approach. We assume that the number of traps is significantly larger than the number of electrons and that the trap depths \(\omega_a\) are uniformly distributed. The thermal energy at \(T = 3.4\,\mathrm{K}\) is \(k_B T \approx 70\,\mathrm{GHz}\cdot h\). Electrons trapped in deeper potential wells (\(\hbar \omega_a \gg k_B T\)) remain trapped longer, whereas those in shallower traps escape more quickly. Following previous studies on thermal escape time from quantum wells~\cite{Gurioli1992-ms,Swank1963-ap}, we assume that the thermal escape time of an electron from a trap scales as \(\exp(\hbar \omega_a / k_B T)\), and similarly, the probability of an electron occupying a trap at a given \(\omega_a\) follows the same dependence. We calculate and plot \(\Delta f\) and \(Q_\mathrm{e}\) by integrating over trap depths ranging from \(\omega_a/2\pi = 0\) to \(200\,\mathrm{GHz}\). As shown in Fig.~\ref{fig:f_Q_Sim}(c), this model successfully reproduces the experimentally observed behavior where \(Q_\mathrm{int}\) does not degrade at \(\Delta f = -0.9\%\), yielding \(1/Q_\mathrm{e} = 3.7\times10^{-4}\). Note that the effective electron density \(n_e\) in this model refers only to electrons trapped within this frequency range, estimated to be approximately \(9.6\times10^{9}\)\,$\mathrm{cm}^{-2}$ for \(\Delta f = -0.9\%\). Electrons trapped deeper than \(200\,\mathrm{GHz}\) do not significantly contribute to \(\Delta f\) or \(Q_\mathrm{int}\) within this framework as seen in Fig.~\ref{fig:f_Q_Sim}(a,b).

\section{Power dependence of $Q_\mathrm{int}$}
\label{Power_Qint}

Figure~\ref{fig:Qint_vs_nph} shows \( Q_{\mathrm{int}} \) and \( Q_{\mathrm{ext}} \) of Resonators~1 and 2 as a function of the average photon number \( \langle n_{\mathrm{ph}} \rangle \) in the resonator, measured for the bare resonator, and with neon and electrons present. Since this measurement was performed at $10\,\mathrm{mK}$, we assume that the effect of electrons is minimal under our surface conditions. As shown in Fig.~\ref{fig:e_deposition}, the quality factor does not degrade upon electron loading. Thus, the observed variations in \( Q_{\mathrm{int}} \) are primarily associated with the presence or absence of neon. The solid lines in Fig.~\ref{fig:Qint_vs_nph} represent fitting curves based on the following TLS loss model:

\begin{equation}
  \frac{1}{Q_\mathrm{int}} = \frac{F}{Q_\mathrm{TLS,0}\sqrt{1 + \left( \frac{\langle n_{\mathrm{ph}} \rangle}{n_{\mathrm{sat}}} \right)^{\beta}}} + \frac{1}{Q_{\mathrm{other}}},
    \label{eq:TLS_loss}
\end{equation}
where \( F \) is the participation ratio of the electric energy in the region where the TLS exists, \( 1/Q_{\mathrm{TLS},0} \) is the intrinsic TLS loss in the zero-photon and zero-temperature limit, \( n_{\mathrm{sat}} \) is the saturation photon number, and \( \beta \) is an empirical parameter that accounts for TLS dynamics. Since the data points with lower \( \langle n_{\mathrm{ph}} \rangle \) have a lower signal-to-noise ratio (SNR), we set the fitting weight \( w_i \) of each data point \( i \) proportional to its SNR. The fitting results demonstrate a good fit and suggest that \( Q_{\mathrm{int}} \) of the NbTiN resonators is limited by two-level system (TLS) loss ~\cite{Gao2008-ox,muller2022magnetic,bruno2015reducing,barends2010minimal}. With neon and electrons, \( Q_{\mathrm{TLS},0}/F \) increases by a factor of \(\sim3\) compared to the bare resonator. This is likely due to the gap between the resonator and the ground being filled with neon, which has a higher dielectric constant than vacuum~\cite{Lane2020-pc}. As a result, the participation ratio \( F \) decreases~\cite{Gao2008-ox}. Additionally, \( n_\mathrm{sat} \) also increases, indicating that more microwave power is required to saturate the TLS due to the presence of neon. However, the observed changes are larger than expected based on neon’s relative permittivity of 1.244. This suggests that some other mechanism might also be involved, and further study is needed.

\section{Two-tone measurement}
\label{2_tone}
\begin{figure}[H]
    \centering
    \includegraphics[width=\linewidth]{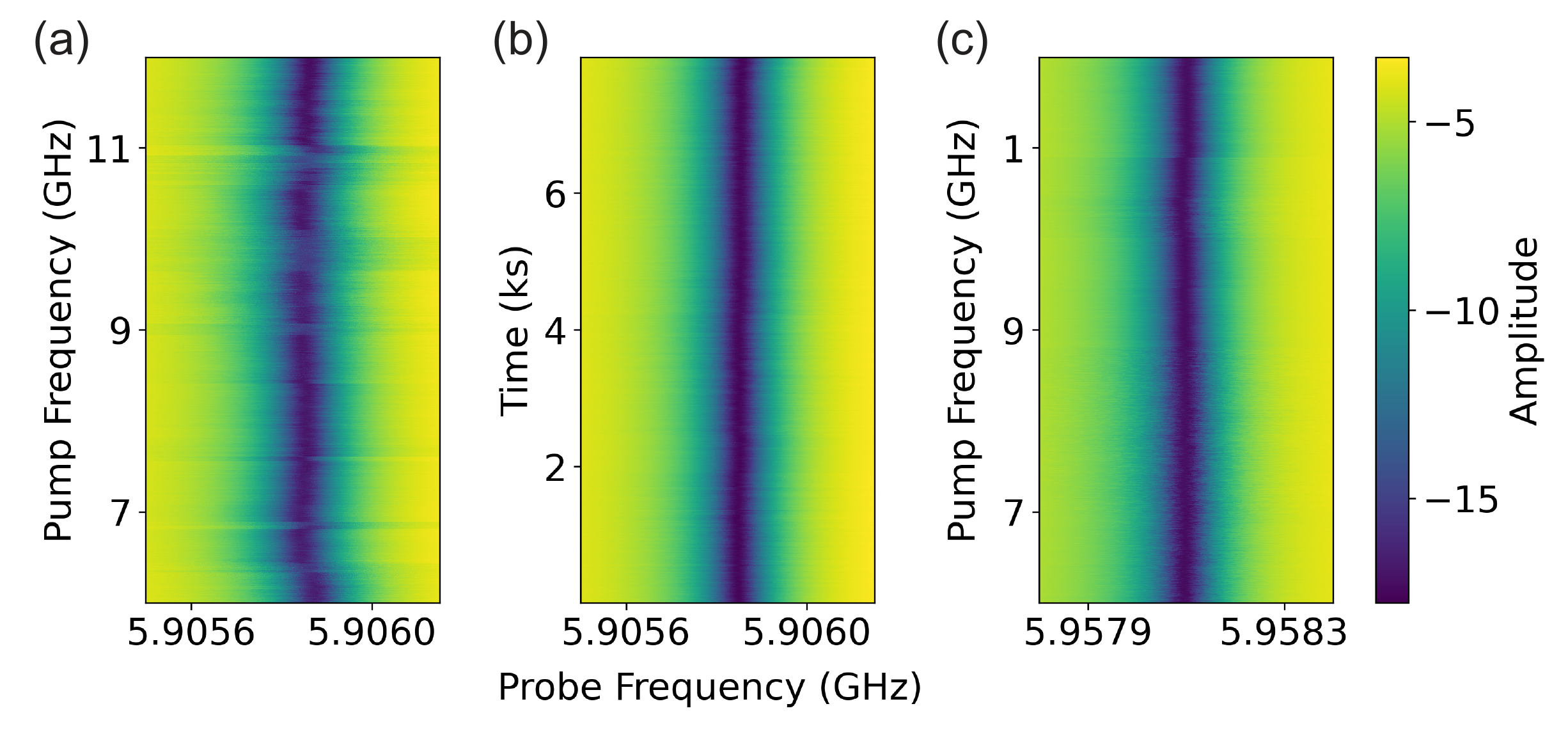}
    \caption{
       Resonance peaks of Resonator 2 measured at $10\,\mathrm{mK}$. We observed that the resonance frequency changed only when both neon and electrons were present. (a) Two-tone measurement with both neon and electrons. 
(b) Stability measurement of the resonance peak over time. (c) Two-tone measurement following the removal of electrons and neon.}
    \label{fig:two-tone}
\end{figure}

Figure~\ref{fig:two-tone} shows the two-tone measurement on Resonator~2. At room temperature, a probe signal with a frequency near the resonance frequency is combined with a pump signal in the $6-12\,\mathrm{GHz}$ range and injected into port~1. The pump signal excites from the ground orbital state to the orbital excited state when their energy difference corresponds to the pump signal frequency, causing a shift in the resonance frequency. 


\section{Micromagnet Simulation}
\label{Mag_sim}

The numerical calculation of the local magnetic fields in different directions is done by modeling each Co micromagnet as a uniformly magnetized rectangular block. We treat the block as a sum of many small magnetic dipoles and integrate their fields over the block’s length, width, and thickness~\cite{Goldman}. 

In this work, we evaluate two Co blocks with magnetization $M=1.7\,\text{T}$, dimensions $1.5\,\mu\text{m}\times1.5\,\mu\text{m}\times t$, where the $t$ represents for different thickness to be simulated. The magnetic field along the axis of interest was calculated with the Magpylib Python package.

\begin{figure}[H]
    \centering
    \includegraphics[width=\linewidth]{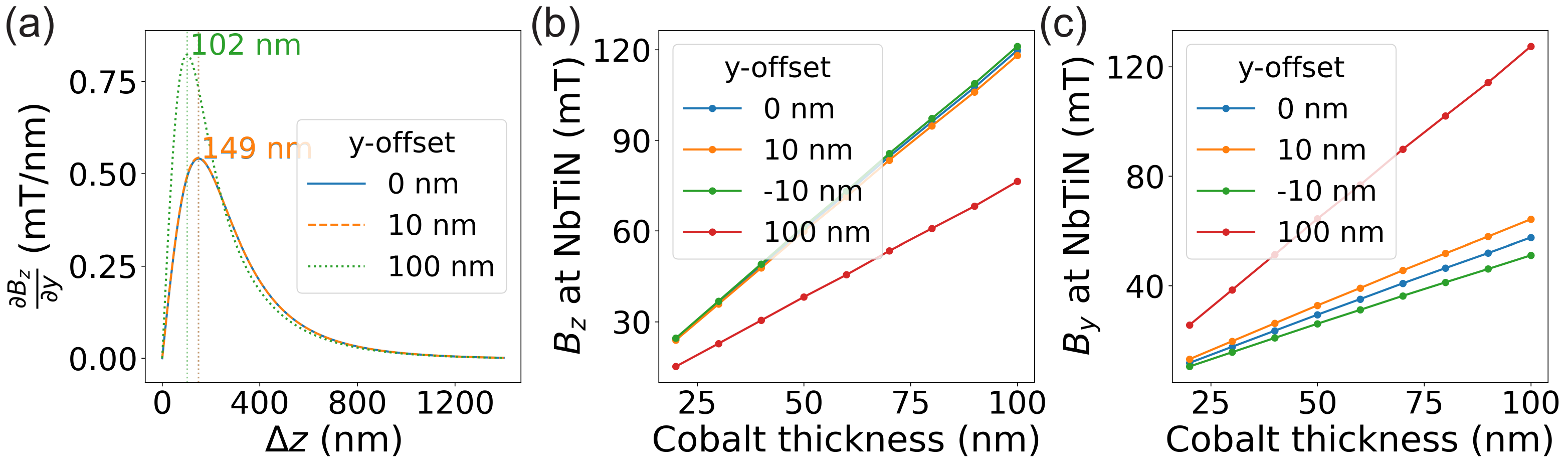}
    \caption{(a) $\frac{\partial B_z}{\partial y}$ as a function of $\Delta z$ for a 100-nm-thick cobalt magnet with misalignment along the $y$-direction. (b) $B_z$ and (c) $B_y$ at the resonator position as functions of cobalt thickness, both with misalignment along the $y$-direction.}
    \label{fig:mag_sim}
\end{figure}

As shown in Fig.~\ref{fig:micro_magnet}(b), $d B_z/d y$ is maximized for $\Delta z \approx 146$~nm. We also simulate the displacement sensitivity. A 10 nm shift of the micromagnets along $y$ changes $dB_z/dy$ only slightly (orange dashed line in Fig.~\ref{fig:mag_sim}(a)), this indicates small misalignments have little effect. A larger 100 nm shift (green dotted line) increases $dB_z/dy$ at the same $\Delta z$. Thus, within normal fabrication tolerances, the gradient—and therefore the coupling strength—can be maintained or even improved as the electron moves closer to one of the magnets.

The absolute magnetic fields with misalignment along the $y$-direction were also simulated (Fig.~\ref{fig:mag_sim}(b,c)). The fields at the resonator position stay well below the NbTiN critical field. The field at the resonator position remains well below the critical field of NbTiN. Therefore, the micromagnets are not expected to degrade the resonator’s quality factor. In addition, we note that displacement along the $x$-direction has a negligible impact on both the magnetic field and its gradient.

\section{Kinetic inductance}\label{K_in}
From separate measurements at room temperature, we estimated the magnetic penetration depth at zero temperature for the NbTiN film with a thickness of \( D = 20\,\mathrm{nm} \), used in the experiments, to be \( \lambda = 390\,\mathrm{nm} \). The inductance per square \( L_{\square} \) is calculated as \(
L_\square = \frac{\mu_0 \lambda^2}{D} =9.6 \, \mathrm{pH}.
\)
The total inductance of the resonator is determined by multiplying \( L_\square \) by the number of squares that fit within the geometry. Thus, the kinetic inductance of Resonator~1 is calculated as \(
L = \frac{L_\square l}{w} = 9.6\, \mathrm{pH} \cdot \frac{1.45 \, \mathrm{mm}}{100 \, \mathrm{nm}} = 139 \, \mathrm{nH},
\) where $l$ and $w$ are the length and the width of the resonator, respectively. As shown in Table~\ref{reso_table}, the resonance frequencies obtained from both simulation and experiment are in good agreement for Resonator 1, thereby validating this inductance estimation.




\bibliography{library}

\begin{thebibliography}{58}%
\makeatletter
\providecommand \@ifxundefined [1]{%
 \@ifx{#1\undefined}
}%
\providecommand \@ifnum [1]{%
 \ifnum #1\expandafter \@firstoftwo
 \else \expandafter \@secondoftwo
 \fi
}%
\providecommand \@ifx [1]{%
 \ifx #1\expandafter \@firstoftwo
 \else \expandafter \@secondoftwo
 \fi
}%
\providecommand \natexlab [1]{#1}%
\providecommand \enquote  [1]{``#1''}%
\providecommand \bibnamefont  [1]{#1}%
\providecommand \bibfnamefont [1]{#1}%
\providecommand \citenamefont [1]{#1}%
\providecommand \href@noop [0]{\@secondoftwo}%
\providecommand \href [0]{\begingroup \@sanitize@url \@href}%
\providecommand \@href[1]{\@@startlink{#1}\@@href}%
\providecommand \@@href[1]{\endgroup#1\@@endlink}%
\providecommand \@sanitize@url [0]{\catcode `\\12\catcode `\$12\catcode `\&12\catcode `\#12\catcode `\^12\catcode `\_12\catcode `\%12\relax}%
\providecommand \@@startlink[1]{}%
\providecommand \@@endlink[0]{}%
\providecommand \url  [0]{\begingroup\@sanitize@url \@url }%
\providecommand \@url [1]{\endgroup\@href {#1}{\urlprefix }}%
\providecommand \urlprefix  [0]{URL }%
\providecommand \Eprint [0]{\href }%
\providecommand \doibase [0]{https://doi.org/}%
\providecommand \selectlanguage [0]{\@gobble}%
\providecommand \bibinfo  [0]{\@secondoftwo}%
\providecommand \bibfield  [0]{\@secondoftwo}%
\providecommand \translation [1]{[#1]}%
\providecommand \BibitemOpen [0]{}%
\providecommand \bibitemStop [0]{}%
\providecommand \bibitemNoStop [0]{.\EOS\space}%
\providecommand \EOS [0]{\spacefactor3000\relax}%
\providecommand \BibitemShut  [1]{\csname bibitem#1\endcsname}%
\let\auto@bib@innerbib\@empty
\bibitem [{\citenamefont {Petersson}\ \emph {et~al.}(2012)\citenamefont {Petersson}, \citenamefont {McFaul}, \citenamefont {Schroer}, \citenamefont {Jung}, \citenamefont {Taylor}, \citenamefont {Houck},\ and\ \citenamefont {Petta}}]{Petersson2012}%
  \BibitemOpen
  \bibfield  {author} {\bibinfo {author} {\bibfnamefont {K.~D.}\ \bibnamefont {Petersson}}, \bibinfo {author} {\bibfnamefont {L.~W.}\ \bibnamefont {McFaul}}, \bibinfo {author} {\bibfnamefont {M.~D.}\ \bibnamefont {Schroer}}, \bibinfo {author} {\bibfnamefont {M.}~\bibnamefont {Jung}}, \bibinfo {author} {\bibfnamefont {J.~M.}\ \bibnamefont {Taylor}}, \bibinfo {author} {\bibfnamefont {A.~A.}\ \bibnamefont {Houck}},\ and\ \bibinfo {author} {\bibfnamefont {J.~R.}\ \bibnamefont {Petta}},\ }\bibfield  {title} {\bibinfo {title} {{Circuit quantum electrodynamics with a spin qubit.}},\ }\href {https://doi.org/10.1038/nature11559} {\bibfield  {journal} {\bibinfo  {journal} {Nature}\ }\textbf {\bibinfo {volume} {490}},\ \bibinfo {pages} {380} (\bibinfo {year} {2012})}\BibitemShut {NoStop}%
\bibitem [{\citenamefont {Mi}\ \emph {et~al.}(2017{\natexlab{a}})\citenamefont {Mi}, \citenamefont {Cady}, \citenamefont {Zajac}, \citenamefont {Deelman},\ and\ \citenamefont {Petta}}]{Mi2017-sp}%
  \BibitemOpen
  \bibfield  {author} {\bibinfo {author} {\bibfnamefont {X.}~\bibnamefont {Mi}}, \bibinfo {author} {\bibfnamefont {J.~V.}\ \bibnamefont {Cady}}, \bibinfo {author} {\bibfnamefont {D.~M.}\ \bibnamefont {Zajac}}, \bibinfo {author} {\bibfnamefont {P.~W.}\ \bibnamefont {Deelman}},\ and\ \bibinfo {author} {\bibfnamefont {J.~R.}\ \bibnamefont {Petta}},\ }\bibfield  {title} {\bibinfo {title} {Strong coupling of a single electron in silicon to a microwave photon},\ }\href@noop {} {\bibfield  {journal} {\bibinfo  {journal} {Science}\ }\textbf {\bibinfo {volume} {355}},\ \bibinfo {pages} {156} (\bibinfo {year} {2017}{\natexlab{a}})}\BibitemShut {NoStop}%
\bibitem [{\citenamefont {Stockklauser}\ \emph {et~al.}(2017)\citenamefont {Stockklauser}, \citenamefont {Scarlino}, \citenamefont {Koski}, \citenamefont {Gasparinetti}, \citenamefont {Andersen}, \citenamefont {Reichl}, \citenamefont {Wegscheider}, \citenamefont {Ihn}, \citenamefont {Ensslin},\ and\ \citenamefont {Wallraff}}]{Stockklauser2017-ao}%
  \BibitemOpen
  \bibfield  {author} {\bibinfo {author} {\bibfnamefont {A.}~\bibnamefont {Stockklauser}}, \bibinfo {author} {\bibfnamefont {P.}~\bibnamefont {Scarlino}}, \bibinfo {author} {\bibfnamefont {J.~V.}\ \bibnamefont {Koski}}, \bibinfo {author} {\bibfnamefont {S.}~\bibnamefont {Gasparinetti}}, \bibinfo {author} {\bibfnamefont {C.~K.}\ \bibnamefont {Andersen}}, \bibinfo {author} {\bibfnamefont {C.}~\bibnamefont {Reichl}}, \bibinfo {author} {\bibfnamefont {W.}~\bibnamefont {Wegscheider}}, \bibinfo {author} {\bibfnamefont {T.}~\bibnamefont {Ihn}}, \bibinfo {author} {\bibfnamefont {K.}~\bibnamefont {Ensslin}},\ and\ \bibinfo {author} {\bibfnamefont {A.}~\bibnamefont {Wallraff}},\ }\bibfield  {title} {\bibinfo {title} {Strong coupling cavity qed with gate-defined double quantum dots enabled by a high impedance resonator},\ }\href@noop {} {\bibfield  {journal} {\bibinfo  {journal} {Phys. Rev. X}\ }\textbf {\bibinfo {volume} {7}},\ \bibinfo {pages} {011030} (\bibinfo {year} {2017})}\BibitemShut {NoStop}%
\bibitem [{\citenamefont {Landig}\ \emph {et~al.}(2018)\citenamefont {Landig}, \citenamefont {Koski}, \citenamefont {Scarlino}, \citenamefont {Mendes}, \citenamefont {Blais}, \citenamefont {Reichl}, \citenamefont {Wegscheider}, \citenamefont {Wallraff}, \citenamefont {Ensslin},\ and\ \citenamefont {Ihn}}]{Landig2018-wt}%
  \BibitemOpen
  \bibfield  {author} {\bibinfo {author} {\bibfnamefont {A.~J.}\ \bibnamefont {Landig}}, \bibinfo {author} {\bibfnamefont {J.~V.}\ \bibnamefont {Koski}}, \bibinfo {author} {\bibfnamefont {P.}~\bibnamefont {Scarlino}}, \bibinfo {author} {\bibfnamefont {U.~C.}\ \bibnamefont {Mendes}}, \bibinfo {author} {\bibfnamefont {A.}~\bibnamefont {Blais}}, \bibinfo {author} {\bibfnamefont {C.}~\bibnamefont {Reichl}}, \bibinfo {author} {\bibfnamefont {W.}~\bibnamefont {Wegscheider}}, \bibinfo {author} {\bibfnamefont {A.}~\bibnamefont {Wallraff}}, \bibinfo {author} {\bibfnamefont {K.}~\bibnamefont {Ensslin}},\ and\ \bibinfo {author} {\bibfnamefont {T.}~\bibnamefont {Ihn}},\ }\bibfield  {title} {\bibinfo {title} {Coherent spin-photon coupling using a resonant exchange qubit},\ }\href@noop {} {\bibfield  {journal} {\bibinfo  {journal} {Nature}\ }\textbf {\bibinfo {volume} {560}},\ \bibinfo {pages} {179} (\bibinfo {year} {2018})}\BibitemShut {NoStop}%
\bibitem [{\citenamefont {Samkharadze}\ \emph {et~al.}(2016)\citenamefont {Samkharadze}, \citenamefont {Bruno}, \citenamefont {Scarlino}, \citenamefont {Zheng}, \citenamefont {DiVincenzo}, \citenamefont {DiCarlo},\ and\ \citenamefont {Vandersypen}}]{Samkharadze2016-xh}%
  \BibitemOpen
  \bibfield  {author} {\bibinfo {author} {\bibfnamefont {N.}~\bibnamefont {Samkharadze}}, \bibinfo {author} {\bibfnamefont {A.}~\bibnamefont {Bruno}}, \bibinfo {author} {\bibfnamefont {P.}~\bibnamefont {Scarlino}}, \bibinfo {author} {\bibfnamefont {G.}~\bibnamefont {Zheng}}, \bibinfo {author} {\bibfnamefont {D.~P.}\ \bibnamefont {DiVincenzo}}, \bibinfo {author} {\bibfnamefont {L.}~\bibnamefont {DiCarlo}},\ and\ \bibinfo {author} {\bibfnamefont {L.~M.~K.}\ \bibnamefont {Vandersypen}},\ }\bibfield  {title} {\bibinfo {title} {High-kinetic-inductance superconducting nanowire resonators for circuit qed in a magnetic field},\ }\href@noop {} {\bibfield  {journal} {\bibinfo  {journal} {Phys. Rev. Appl.}\ }\textbf {\bibinfo {volume} {5}},\ \bibinfo {pages} {044004} (\bibinfo {year} {2016})}\BibitemShut {NoStop}%
\bibitem [{\citenamefont {Mi}\ \emph {et~al.}(2018)\citenamefont {Mi}, \citenamefont {Benito}, \citenamefont {Putz}, \citenamefont {Zajac}, \citenamefont {Taylor}, \citenamefont {Burkard},\ and\ \citenamefont {Petta}}]{Mi2018-co}%
  \BibitemOpen
  \bibfield  {author} {\bibinfo {author} {\bibfnamefont {X.}~\bibnamefont {Mi}}, \bibinfo {author} {\bibfnamefont {M.}~\bibnamefont {Benito}}, \bibinfo {author} {\bibfnamefont {S.}~\bibnamefont {Putz}}, \bibinfo {author} {\bibfnamefont {D.~M.}\ \bibnamefont {Zajac}}, \bibinfo {author} {\bibfnamefont {J.~M.}\ \bibnamefont {Taylor}}, \bibinfo {author} {\bibfnamefont {G.}~\bibnamefont {Burkard}},\ and\ \bibinfo {author} {\bibfnamefont {J.~R.}\ \bibnamefont {Petta}},\ }\bibfield  {title} {\bibinfo {title} {A coherent spin-photon interface in silicon},\ }\href@noop {} {\bibfield  {journal} {\bibinfo  {journal} {Nature}\ }\textbf {\bibinfo {volume} {555}},\ \bibinfo {pages} {599} (\bibinfo {year} {2018})}\BibitemShut {NoStop}%
\bibitem [{\citenamefont {Borjans}\ \emph {et~al.}(2020)\citenamefont {Borjans}, \citenamefont {Croot}, \citenamefont {Mi}, \citenamefont {Gullans},\ and\ \citenamefont {Petta}}]{Borjans2020-do}%
  \BibitemOpen
  \bibfield  {author} {\bibinfo {author} {\bibfnamefont {F.}~\bibnamefont {Borjans}}, \bibinfo {author} {\bibfnamefont {X.~G.}\ \bibnamefont {Croot}}, \bibinfo {author} {\bibfnamefont {X.}~\bibnamefont {Mi}}, \bibinfo {author} {\bibfnamefont {M.~J.}\ \bibnamefont {Gullans}},\ and\ \bibinfo {author} {\bibfnamefont {J.~R.}\ \bibnamefont {Petta}},\ }\bibfield  {title} {\bibinfo {title} {Resonant microwave-mediated interactions between distant electron spins},\ }\href@noop {} {\bibfield  {journal} {\bibinfo  {journal} {Nature}\ }\textbf {\bibinfo {volume} {577}},\ \bibinfo {pages} {195} (\bibinfo {year} {2020})}\BibitemShut {NoStop}%
\bibitem [{\citenamefont {Bøttcher}\ \emph {et~al.}(2022)\citenamefont {Bøttcher}, \citenamefont {Harvey}, \citenamefont {Fallahi}, \citenamefont {Gardner}, \citenamefont {Manfra}, \citenamefont {Vool}, \citenamefont {Bartlett},\ and\ \citenamefont {Yacoby}}]{Bottcher2022-dp}%
  \BibitemOpen
  \bibfield  {author} {\bibinfo {author} {\bibfnamefont {C.~G.~L.}\ \bibnamefont {Bøttcher}}, \bibinfo {author} {\bibfnamefont {S.~P.}\ \bibnamefont {Harvey}}, \bibinfo {author} {\bibfnamefont {S.}~\bibnamefont {Fallahi}}, \bibinfo {author} {\bibfnamefont {G.~C.}\ \bibnamefont {Gardner}}, \bibinfo {author} {\bibfnamefont {M.~J.}\ \bibnamefont {Manfra}}, \bibinfo {author} {\bibfnamefont {U.}~\bibnamefont {Vool}}, \bibinfo {author} {\bibfnamefont {S.~D.}\ \bibnamefont {Bartlett}},\ and\ \bibinfo {author} {\bibfnamefont {A.}~\bibnamefont {Yacoby}},\ }\bibfield  {title} {\bibinfo {title} {Parametric longitudinal coupling between a high-impedance superconducting resonator and a semiconductor quantum dot singlet-triplet spin qubit},\ }\href@noop {} {\bibfield  {journal} {\bibinfo  {journal} {Nat. Commun.}\ }\textbf {\bibinfo {volume} {13}},\ \bibinfo {pages} {4773} (\bibinfo {year} {2022})}\BibitemShut {NoStop}%
\bibitem [{\citenamefont {Harvey-Collard}\ \emph {et~al.}(2022)\citenamefont {Harvey-Collard}, \citenamefont {Dijkema}, \citenamefont {Zheng}, \citenamefont {Sammak}, \citenamefont {Scappucci},\ and\ \citenamefont {Vandersypen}}]{Harvey-Collard2022-jh}%
  \BibitemOpen
  \bibfield  {author} {\bibinfo {author} {\bibfnamefont {P.}~\bibnamefont {Harvey-Collard}}, \bibinfo {author} {\bibfnamefont {J.}~\bibnamefont {Dijkema}}, \bibinfo {author} {\bibfnamefont {G.}~\bibnamefont {Zheng}}, \bibinfo {author} {\bibfnamefont {A.}~\bibnamefont {Sammak}}, \bibinfo {author} {\bibfnamefont {G.}~\bibnamefont {Scappucci}},\ and\ \bibinfo {author} {\bibfnamefont {L.~M.~K.}\ \bibnamefont {Vandersypen}},\ }\bibfield  {title} {\bibinfo {title} {Coherent spin-spin coupling mediated by virtual microwave photons},\ }\href@noop {} {\bibfield  {journal} {\bibinfo  {journal} {Phys. Rev. X}\ }\textbf {\bibinfo {volume} {12}},\ \bibinfo {pages} {021026} (\bibinfo {year} {2022})}\BibitemShut {NoStop}%
\bibitem [{\citenamefont {Dijkema}\ \emph {et~al.}(2025)\citenamefont {Dijkema}, \citenamefont {Xue}, \citenamefont {Harvey-Collard} \emph {et~al.}}]{Dijkema2025}%
  \BibitemOpen
  \bibfield  {author} {\bibinfo {author} {\bibfnamefont {J.}~\bibnamefont {Dijkema}}, \bibinfo {author} {\bibfnamefont {X.}~\bibnamefont {Xue}}, \bibinfo {author} {\bibfnamefont {P.}~\bibnamefont {Harvey-Collard}}, \emph {et~al.},\ }\bibfield  {title} {\bibinfo {title} {Cavity-mediated iswap oscillations between distant spins},\ }\href {https://doi.org/10.1038/s41567-024-02694-8} {\bibfield  {journal} {\bibinfo  {journal} {Nat. Phys.}\ }\textbf {\bibinfo {volume} {21}},\ \bibinfo {pages} {168} (\bibinfo {year} {2025})}\BibitemShut {NoStop}%
\bibitem [{\citenamefont {Zhou}\ \emph {et~al.}(2022)\citenamefont {Zhou}, \citenamefont {Koolstra}, \citenamefont {Zhang}, \citenamefont {Yang}, \citenamefont {Han}, \citenamefont {Dizdar}, \citenamefont {Li}, \citenamefont {Divan}, \citenamefont {Guo}, \citenamefont {Murch}, \citenamefont {Schuster},\ and\ \citenamefont {Jin}}]{Zhou2022-nk}%
  \BibitemOpen
  \bibfield  {author} {\bibinfo {author} {\bibfnamefont {X.}~\bibnamefont {Zhou}}, \bibinfo {author} {\bibfnamefont {G.}~\bibnamefont {Koolstra}}, \bibinfo {author} {\bibfnamefont {X.}~\bibnamefont {Zhang}}, \bibinfo {author} {\bibfnamefont {G.}~\bibnamefont {Yang}}, \bibinfo {author} {\bibfnamefont {X.}~\bibnamefont {Han}}, \bibinfo {author} {\bibfnamefont {B.}~\bibnamefont {Dizdar}}, \bibinfo {author} {\bibfnamefont {X.}~\bibnamefont {Li}}, \bibinfo {author} {\bibfnamefont {R.}~\bibnamefont {Divan}}, \bibinfo {author} {\bibfnamefont {W.}~\bibnamefont {Guo}}, \bibinfo {author} {\bibfnamefont {K.~W.}\ \bibnamefont {Murch}}, \bibinfo {author} {\bibfnamefont {D.~I.}\ \bibnamefont {Schuster}},\ and\ \bibinfo {author} {\bibfnamefont {D.}~\bibnamefont {Jin}},\ }\bibfield  {title} {\bibinfo {title} {Single electrons on solid neon as a solid-state qubit platform},\ }\href@noop {} {\bibfield  {journal} {\bibinfo  {journal} {Nature}\ }\textbf {\bibinfo {volume} {605}},\ \bibinfo {pages} {46} (\bibinfo {year}
  {2022})}\BibitemShut {NoStop}%
\bibitem [{\citenamefont {Zhou}\ \emph {et~al.}(2024)\citenamefont {Zhou}, \citenamefont {Li}, \citenamefont {Chen}, \citenamefont {Koolstra}, \citenamefont {Yang}, \citenamefont {Dizdar}, \citenamefont {Huang}, \citenamefont {Wang}, \citenamefont {Han}, \citenamefont {Zhang}, \citenamefont {Schuster},\ and\ \citenamefont {Jin}}]{Zhou2023-iw}%
  \BibitemOpen
  \bibfield  {author} {\bibinfo {author} {\bibfnamefont {X.}~\bibnamefont {Zhou}}, \bibinfo {author} {\bibfnamefont {X.}~\bibnamefont {Li}}, \bibinfo {author} {\bibfnamefont {Q.}~\bibnamefont {Chen}}, \bibinfo {author} {\bibfnamefont {G.}~\bibnamefont {Koolstra}}, \bibinfo {author} {\bibfnamefont {G.}~\bibnamefont {Yang}}, \bibinfo {author} {\bibfnamefont {B.}~\bibnamefont {Dizdar}}, \bibinfo {author} {\bibfnamefont {Y.}~\bibnamefont {Huang}}, \bibinfo {author} {\bibfnamefont {C.~S.}\ \bibnamefont {Wang}}, \bibinfo {author} {\bibfnamefont {X.}~\bibnamefont {Han}}, \bibinfo {author} {\bibfnamefont {X.}~\bibnamefont {Zhang}}, \bibinfo {author} {\bibfnamefont {D.~I.}\ \bibnamefont {Schuster}},\ and\ \bibinfo {author} {\bibfnamefont {D.}~\bibnamefont {Jin}},\ }\bibfield  {title} {\bibinfo {title} {Electron charge qubit with 0.1 millisecond coherence time},\ }\href@noop {} {\bibfield  {journal} {\bibinfo  {journal} {Nat. Phys.}\ }\textbf {\bibinfo {volume} {20}},\ \bibinfo {pages} {116} (\bibinfo {year}
  {2024})}\BibitemShut {NoStop}%
\bibitem [{\citenamefont {Li}\ \emph {et~al.}(2025{\natexlab{a}})\citenamefont {Li}, \citenamefont {Wang}, \citenamefont {Dizdar}, \citenamefont {Huang}, \citenamefont {Wen}, \citenamefont {Guo}, \citenamefont {Zhang}, \citenamefont {Han}, \citenamefont {Zhou},\ and\ \citenamefont {Jin}}]{Li2025-em}%
  \BibitemOpen
  \bibfield  {author} {\bibinfo {author} {\bibfnamefont {X.}~\bibnamefont {Li}}, \bibinfo {author} {\bibfnamefont {C.~S.}\ \bibnamefont {Wang}}, \bibinfo {author} {\bibfnamefont {B.}~\bibnamefont {Dizdar}}, \bibinfo {author} {\bibfnamefont {Y.}~\bibnamefont {Huang}}, \bibinfo {author} {\bibfnamefont {Y.}~\bibnamefont {Wen}}, \bibinfo {author} {\bibfnamefont {W.}~\bibnamefont {Guo}}, \bibinfo {author} {\bibfnamefont {X.}~\bibnamefont {Zhang}}, \bibinfo {author} {\bibfnamefont {X.}~\bibnamefont {Han}}, \bibinfo {author} {\bibfnamefont {X.}~\bibnamefont {Zhou}},\ and\ \bibinfo {author} {\bibfnamefont {D.}~\bibnamefont {Jin}},\ }\bibfield  {title} {\bibinfo {title} {Noise-resilient host for electron qubit operation up to 0.4 k},\ }\href {https://arxiv.org/abs/2502.01005} {\bibfield  {journal} {\bibinfo  {journal} {arXiv preprint}\ } (\bibinfo {year} {2025}{\natexlab{a}})},\ \Eprint {https://arxiv.org/abs/2502.01005} {arXiv:2502.01005} \BibitemShut {NoStop}%
\bibitem [{\citenamefont {Li}\ \emph {et~al.}(2025{\natexlab{b}})\citenamefont {Li}, \citenamefont {Zou}, \citenamefont {Chen},\ and\ \citenamefont {Jin}}]{li2025electron}%
  \BibitemOpen
  \bibfield  {author} {\bibinfo {author} {\bibfnamefont {X.}~\bibnamefont {Li}}, \bibinfo {author} {\bibfnamefont {S.}~\bibnamefont {Zou}}, \bibinfo {author} {\bibfnamefont {Q.}~\bibnamefont {Chen}},\ and\ \bibinfo {author} {\bibfnamefont {D.}~\bibnamefont {Jin}},\ }\bibfield  {title} {\bibinfo {title} {Electron charge coherence on a solid neon surface},\ }\href {https://arxiv.org/abs/2507.20476} {\bibfield  {journal} {\bibinfo  {journal} {arXiv preprint}\ } (\bibinfo {year} {2025}{\natexlab{b}})},\ \Eprint {https://arxiv.org/abs/2507.20476} {arXiv:2507.20476} \BibitemShut {NoStop}%
\bibitem [{\citenamefont {Chen}\ \emph {et~al.}(2022)\citenamefont {Chen}, \citenamefont {Martin}, \citenamefont {Jiang},\ and\ \citenamefont {Jin}}]{Chen2022-on}%
  \BibitemOpen
  \bibfield  {author} {\bibinfo {author} {\bibfnamefont {Q.}~\bibnamefont {Chen}}, \bibinfo {author} {\bibfnamefont {I.}~\bibnamefont {Martin}}, \bibinfo {author} {\bibfnamefont {L.}~\bibnamefont {Jiang}},\ and\ \bibinfo {author} {\bibfnamefont {D.}~\bibnamefont {Jin}},\ }\bibfield  {title} {\bibinfo {title} {Electron spin coherence on a solid neon surface},\ }\href@noop {} {\bibfield  {journal} {\bibinfo  {journal} {Quantum Sci. Technol.}\ }\textbf {\bibinfo {volume} {7}},\ \bibinfo {pages} {045016} (\bibinfo {year} {2022})}\BibitemShut {NoStop}%
\bibitem [{\citenamefont {Benito}\ \emph {et~al.}(2017)\citenamefont {Benito}, \citenamefont {Mi}, \citenamefont {Taylor}, \citenamefont {Petta},\ and\ \citenamefont {Burkard}}]{Benito2017-ok}%
  \BibitemOpen
  \bibfield  {author} {\bibinfo {author} {\bibfnamefont {M.}~\bibnamefont {Benito}}, \bibinfo {author} {\bibfnamefont {X.}~\bibnamefont {Mi}}, \bibinfo {author} {\bibfnamefont {J.~M.}\ \bibnamefont {Taylor}}, \bibinfo {author} {\bibfnamefont {J.~R.}\ \bibnamefont {Petta}},\ and\ \bibinfo {author} {\bibfnamefont {G.}~\bibnamefont {Burkard}},\ }\bibfield  {title} {\bibinfo {title} {Input-output theory for spin-photon coupling in si double quantum dots},\ }\href@noop {} {\bibfield  {journal} {\bibinfo  {journal} {Phys. Rev. B}\ }\textbf {\bibinfo {volume} {96}},\ \bibinfo {pages} {235434} (\bibinfo {year} {2017})}\BibitemShut {NoStop}%
\bibitem [{\citenamefont {Benito}\ \emph {et~al.}(2019{\natexlab{a}})\citenamefont {Benito}, \citenamefont {Petta},\ and\ \citenamefont {Burkard}}]{Benito2019-al}%
  \BibitemOpen
  \bibfield  {author} {\bibinfo {author} {\bibfnamefont {M.}~\bibnamefont {Benito}}, \bibinfo {author} {\bibfnamefont {J.~R.}\ \bibnamefont {Petta}},\ and\ \bibinfo {author} {\bibfnamefont {G.}~\bibnamefont {Burkard}},\ }\bibfield  {title} {\bibinfo {title} {Optimized cavity-mediated dispersive two-qubit gates between spin qubits},\ }\href {https://doi.org/10.1103/PhysRevB.100.081412} {\bibfield  {journal} {\bibinfo  {journal} {Phys. Rev. B}\ }\textbf {\bibinfo {volume} {100}},\ \bibinfo {pages} {081412} (\bibinfo {year} {2019}{\natexlab{a}})}\BibitemShut {NoStop}%
\bibitem [{\citenamefont {Samkharadze}\ \emph {et~al.}(2018)\citenamefont {Samkharadze}, \citenamefont {Zheng}, \citenamefont {Kalhor}, \citenamefont {Brousse}, \citenamefont {Sammak}, \citenamefont {Mendes}, \citenamefont {Blais}, \citenamefont {Scappucci},\ and\ \citenamefont {Vandersypen}}]{Samkharadze2018}%
  \BibitemOpen
  \bibfield  {author} {\bibinfo {author} {\bibfnamefont {N.}~\bibnamefont {Samkharadze}}, \bibinfo {author} {\bibfnamefont {G.}~\bibnamefont {Zheng}}, \bibinfo {author} {\bibfnamefont {N.}~\bibnamefont {Kalhor}}, \bibinfo {author} {\bibfnamefont {D.}~\bibnamefont {Brousse}}, \bibinfo {author} {\bibfnamefont {A.}~\bibnamefont {Sammak}}, \bibinfo {author} {\bibfnamefont {U.~C.}\ \bibnamefont {Mendes}}, \bibinfo {author} {\bibfnamefont {A.}~\bibnamefont {Blais}}, \bibinfo {author} {\bibfnamefont {G.}~\bibnamefont {Scappucci}},\ and\ \bibinfo {author} {\bibfnamefont {L.~M.~K.}\ \bibnamefont {Vandersypen}},\ }\bibfield  {title} {\bibinfo {title} {{Strong spin-photon coupling in silicon.}},\ }\href {https://doi.org/10.1126/science.aar4054} {\bibfield  {journal} {\bibinfo  {journal} {Science}\ }\textbf {\bibinfo {volume} {359}},\ \bibinfo {pages} {1123} (\bibinfo {year} {2018})}\BibitemShut {NoStop}%
\bibitem [{\citenamefont {Fowler}\ \emph {et~al.}(2012)\citenamefont {Fowler}, \citenamefont {Mariantoni}, \citenamefont {Martinis},\ and\ \citenamefont {Cleland}}]{Fowler2012}%
  \BibitemOpen
  \bibfield  {author} {\bibinfo {author} {\bibfnamefont {A.~G.}\ \bibnamefont {Fowler}}, \bibinfo {author} {\bibfnamefont {M.}~\bibnamefont {Mariantoni}}, \bibinfo {author} {\bibfnamefont {J.~M.}\ \bibnamefont {Martinis}},\ and\ \bibinfo {author} {\bibfnamefont {A.~N.}\ \bibnamefont {Cleland}},\ }\bibfield  {title} {\bibinfo {title} {{Surface codes: Towards practical large-scale quantum computation}},\ }\href {https://doi.org/10.1103/PhysRevA.86.032324} {\bibfield  {journal} {\bibinfo  {journal} {Phys. Rev. A}\ }\textbf {\bibinfo {volume} {86}},\ \bibinfo {pages} {032324} (\bibinfo {year} {2012})}\BibitemShut {NoStop}%
\bibitem [{\citenamefont {Tokura}\ \emph {et~al.}(2006)\citenamefont {Tokura}, \citenamefont {van~der Wiel}, \citenamefont {Obata},\ and\ \citenamefont {Tarucha}}]{Tokura2006}%
  \BibitemOpen
  \bibfield  {author} {\bibinfo {author} {\bibfnamefont {Y.}~\bibnamefont {Tokura}}, \bibinfo {author} {\bibfnamefont {W.~G.}\ \bibnamefont {van~der Wiel}}, \bibinfo {author} {\bibfnamefont {T.}~\bibnamefont {Obata}},\ and\ \bibinfo {author} {\bibfnamefont {S.}~\bibnamefont {Tarucha}},\ }\bibfield  {title} {\bibinfo {title} {{Coherent Single Electron Spin Control in a Slanting Zeeman Field}},\ }\href {https://doi.org/10.1103/PhysRevLett.96.047202} {\bibfield  {journal} {\bibinfo  {journal} {Phys. Rev. Lett.}\ }\textbf {\bibinfo {volume} {96}},\ \bibinfo {pages} {047202} (\bibinfo {year} {2006})}\BibitemShut {NoStop}%
\bibitem [{\citenamefont {Jennings}\ \emph {et~al.}(2024)\citenamefont {Jennings}, \citenamefont {Zhou}, \citenamefont {Grytsenko},\ and\ \citenamefont {Kawakami}}]{Jennings2024-sb}%
  \BibitemOpen
  \bibfield  {author} {\bibinfo {author} {\bibfnamefont {A.}~\bibnamefont {Jennings}}, \bibinfo {author} {\bibfnamefont {X.}~\bibnamefont {Zhou}}, \bibinfo {author} {\bibfnamefont {I.}~\bibnamefont {Grytsenko}},\ and\ \bibinfo {author} {\bibfnamefont {E.}~\bibnamefont {Kawakami}},\ }\bibfield  {title} {\bibinfo {title} {Quantum computing using floating electrons on cryogenic substrates: Potential and challenges},\ }\href@noop {} {\bibfield  {journal} {\bibinfo  {journal} {Appl. Phys. Lett.}\ }\textbf {\bibinfo {volume} {124}},\ \bibinfo {pages} {120501} (\bibinfo {year} {2024})}\BibitemShut {NoStop}%
\bibitem [{\citenamefont {Kawakami}\ \emph {et~al.}(2023)\citenamefont {Kawakami}, \citenamefont {Chen}, \citenamefont {Benito},\ and\ \citenamefont {Konstantinov}}]{Kawakami2023-vf}%
  \BibitemOpen
  \bibfield  {author} {\bibinfo {author} {\bibfnamefont {E.}~\bibnamefont {Kawakami}}, \bibinfo {author} {\bibfnamefont {J.}~\bibnamefont {Chen}}, \bibinfo {author} {\bibfnamefont {M.}~\bibnamefont {Benito}},\ and\ \bibinfo {author} {\bibfnamefont {D.}~\bibnamefont {Konstantinov}},\ }\bibfield  {title} {\bibinfo {title} {Blueprint for quantum computing using electrons on helium},\ }\href {https://doi.org/10.1103/PhysRevApplied.20.054022} {\bibfield  {journal} {\bibinfo  {journal} {Phys. Rev. Appl.}\ }\textbf {\bibinfo {volume} {20}},\ \bibinfo {pages} {054022} (\bibinfo {year} {2023})}\BibitemShut {NoStop}%
\bibitem [{\citenamefont {Schuster}\ \emph {et~al.}(2010)\citenamefont {Schuster}, \citenamefont {Fragner}, \citenamefont {Dykman}, \citenamefont {Lyon},\ and\ \citenamefont {Schoelkopf}}]{Schuster2010}%
  \BibitemOpen
  \bibfield  {author} {\bibinfo {author} {\bibfnamefont {D.~I.}\ \bibnamefont {Schuster}}, \bibinfo {author} {\bibfnamefont {A.}~\bibnamefont {Fragner}}, \bibinfo {author} {\bibfnamefont {M.~I.}\ \bibnamefont {Dykman}}, \bibinfo {author} {\bibfnamefont {S.~A.}\ \bibnamefont {Lyon}},\ and\ \bibinfo {author} {\bibfnamefont {R.~J.}\ \bibnamefont {Schoelkopf}},\ }\bibfield  {title} {\bibinfo {title} {{Proposal for Manipulating and Detecting Spin and Orbital States of Trapped Electrons on Helium Using Cavity Quantum Electrodynamics}},\ }\href {https://doi.org/10.1103/PhysRevLett.105.040503} {\bibfield  {journal} {\bibinfo  {journal} {Phys. Rev. Lett.}\ }\textbf {\bibinfo {volume} {105}},\ \bibinfo {pages} {040503} (\bibinfo {year} {2010})}\BibitemShut {NoStop}%
\bibitem [{\citenamefont {Kroll}\ \emph {et~al.}(2019)\citenamefont {Kroll}, \citenamefont {Borsoi}, \citenamefont {van~der Enden}, \citenamefont {Uilhoorn}, \citenamefont {de~Jong}, \citenamefont {Quintero-Pérez}, \citenamefont {van Woerkom}, \citenamefont {Bruno}, \citenamefont {Plissard}, \citenamefont {Car}, \citenamefont {Bakkers}, \citenamefont {Cassidy},\ and\ \citenamefont {Kouwenhoven}}]{Kroll2019-eo}%
  \BibitemOpen
  \bibfield  {author} {\bibinfo {author} {\bibfnamefont {J.~G.}\ \bibnamefont {Kroll}}, \bibinfo {author} {\bibfnamefont {F.}~\bibnamefont {Borsoi}}, \bibinfo {author} {\bibfnamefont {K.~L.}\ \bibnamefont {van~der Enden}}, \bibinfo {author} {\bibfnamefont {W.}~\bibnamefont {Uilhoorn}}, \bibinfo {author} {\bibfnamefont {D.}~\bibnamefont {de~Jong}}, \bibinfo {author} {\bibfnamefont {M.}~\bibnamefont {Quintero-Pérez}}, \bibinfo {author} {\bibfnamefont {D.~J.}\ \bibnamefont {van Woerkom}}, \bibinfo {author} {\bibfnamefont {A.}~\bibnamefont {Bruno}}, \bibinfo {author} {\bibfnamefont {S.~R.}\ \bibnamefont {Plissard}}, \bibinfo {author} {\bibfnamefont {D.}~\bibnamefont {Car}}, \bibinfo {author} {\bibfnamefont {E.~P. A.~M.}\ \bibnamefont {Bakkers}}, \bibinfo {author} {\bibfnamefont {M.~C.}\ \bibnamefont {Cassidy}},\ and\ \bibinfo {author} {\bibfnamefont {L.~P.}\ \bibnamefont {Kouwenhoven}},\ }\bibfield  {title} {\bibinfo {title} {Magnetic-field-resilient superconducting coplanar-waveguide resonators for hybrid
  circuit quantum electrodynamics experiments},\ }\href@noop {} {\bibfield  {journal} {\bibinfo  {journal} {Phys. Rev. Appl.}\ }\textbf {\bibinfo {volume} {11}},\ \bibinfo {pages} {064053} (\bibinfo {year} {2019})}\BibitemShut {NoStop}%
\bibitem [{\citenamefont {Probst}\ \emph {et~al.}(2015)\citenamefont {Probst}, \citenamefont {Song}, \citenamefont {Bushev}, \citenamefont {Ustinov},\ and\ \citenamefont {Weides}}]{Probst2015-gs}%
  \BibitemOpen
  \bibfield  {author} {\bibinfo {author} {\bibfnamefont {S.}~\bibnamefont {Probst}}, \bibinfo {author} {\bibfnamefont {F.~B.}\ \bibnamefont {Song}}, \bibinfo {author} {\bibfnamefont {P.~A.}\ \bibnamefont {Bushev}}, \bibinfo {author} {\bibfnamefont {A.~V.}\ \bibnamefont {Ustinov}},\ and\ \bibinfo {author} {\bibfnamefont {M.}~\bibnamefont {Weides}},\ }\bibfield  {title} {\bibinfo {title} {Efficient and robust analysis of complex scattering data under noise in microwave resonators},\ }\href@noop {} {\bibfield  {journal} {\bibinfo  {journal} {Rev. Sci. Instrum.}\ }\textbf {\bibinfo {volume} {86}},\ \bibinfo {pages} {024706} (\bibinfo {year} {2015})}\BibitemShut {NoStop}%
\bibitem [{\citenamefont {Probst}(2024)}]{Probst2024resonator}%
  \BibitemOpen
  \bibfield  {author} {\bibinfo {author} {\bibfnamefont {S.}~\bibnamefont {Probst}},\ }\href@noop {} {\bibinfo {title} {circle fit}} (\bibinfo {year} {2024}),\ \bibinfo {note} {available online: \url{https://github.com/sebastianprobst/resonator_tools}}\BibitemShut {NoStop}%
\bibitem [{\citenamefont {Leiderer}(2025)}]{leiderer2025surface}%
  \BibitemOpen
  \bibfield  {author} {\bibinfo {author} {\bibfnamefont {P.}~\bibnamefont {Leiderer}},\ }\bibfield  {title} {{\selectlanguage {en}\bibinfo {title} {Surface electrons on solid quantum substrates: A brief review}},\ }\href@noop {} {\bibfield  {journal} {\bibinfo  {journal} {J. Low Temp. Phys.}\ }\textbf {\bibinfo {volume} {219}},\ \bibinfo {pages} {262} (\bibinfo {year} {2025})}\BibitemShut {NoStop}%
\bibitem [{\citenamefont {Migone}\ \emph {et~al.}(1986)\citenamefont {Migone}, \citenamefont {Dash}, \citenamefont {Schick},\ and\ \citenamefont {Vilches}}]{migone1986triple}%
  \BibitemOpen
  \bibfield  {author} {\bibinfo {author} {\bibfnamefont {A.}~\bibnamefont {Migone}}, \bibinfo {author} {\bibfnamefont {J.}~\bibnamefont {Dash}}, \bibinfo {author} {\bibfnamefont {M.}~\bibnamefont {Schick}},\ and\ \bibinfo {author} {\bibfnamefont {O.}~\bibnamefont {Vilches}},\ }\bibfield  {title} {\bibinfo {title} {Triple-point wetting of neon films},\ }\href@noop {} {\bibfield  {journal} {\bibinfo  {journal} {Physical Review B}\ }\textbf {\bibinfo {volume} {34}},\ \bibinfo {pages} {6322} (\bibinfo {year} {1986})}\BibitemShut {NoStop}%
\bibitem [{\citenamefont {Kajita}(1984)}]{Kajita1984-zr}%
  \BibitemOpen
  \bibfield  {author} {\bibinfo {author} {\bibfnamefont {K.}~\bibnamefont {Kajita}},\ }\bibfield  {title} {\bibinfo {title} {A new two-dimensional electron system on the surface of solid neon},\ }\href@noop {} {\bibfield  {journal} {\bibinfo  {journal} {Surf. Sci.}\ }\textbf {\bibinfo {volume} {142}},\ \bibinfo {pages} {86} (\bibinfo {year} {1984})}\BibitemShut {NoStop}%
\bibitem [{\citenamefont {Kajita}(1985)}]{Kajita1985-tz}%
  \BibitemOpen
  \bibfield  {author} {\bibinfo {author} {\bibfnamefont {K.}~\bibnamefont {Kajita}},\ }\bibfield  {title} {\bibinfo {title} {Wigner crystallization of two dimensional electrons formed on the surface of solid neon},\ }\href@noop {} {\bibfield  {journal} {\bibinfo  {journal} {J. Phys. Soc. Jpn.}\ }\textbf {\bibinfo {volume} {54}},\ \bibinfo {pages} {4092} (\bibinfo {year} {1985})}\BibitemShut {NoStop}%
\bibitem [{\citenamefont {Nowack}\ \emph {et~al.}(2007)\citenamefont {Nowack}, \citenamefont {Koppens}, \citenamefont {Nazarov},\ and\ \citenamefont {Vandersypen}}]{Nowack2007}%
  \BibitemOpen
  \bibfield  {author} {\bibinfo {author} {\bibfnamefont {K.~C.}\ \bibnamefont {Nowack}}, \bibinfo {author} {\bibfnamefont {F.~H.~L.}\ \bibnamefont {Koppens}}, \bibinfo {author} {\bibfnamefont {Y.~V.}\ \bibnamefont {Nazarov}},\ and\ \bibinfo {author} {\bibfnamefont {L.~M.~K.}\ \bibnamefont {Vandersypen}},\ }\bibfield  {title} {\bibinfo {title} {{Coherent control of a single electron spin with electric fields.}},\ }\href {https://doi.org/10.1126/science.1148092} {\bibfield  {journal} {\bibinfo  {journal} {Science}\ }\textbf {\bibinfo {volume} {318}},\ \bibinfo {pages} {1430} (\bibinfo {year} {2007})}\BibitemShut {NoStop}%
\bibitem [{\citenamefont {Pioro-Ladri{\`{e}}re}\ \emph {et~al.}(2008)\citenamefont {Pioro-Ladri{\`{e}}re}, \citenamefont {Obata}, \citenamefont {Tokura}, \citenamefont {Shin}, \citenamefont {Kubo}, \citenamefont {Yoshida}, \citenamefont {Taniyama},\ and\ \citenamefont {Tarucha}}]{Pioro-Ladriere2008}%
  \BibitemOpen
  \bibfield  {author} {\bibinfo {author} {\bibfnamefont {M.}~\bibnamefont {Pioro-Ladri{\`{e}}re}}, \bibinfo {author} {\bibfnamefont {T.}~\bibnamefont {Obata}}, \bibinfo {author} {\bibfnamefont {Y.}~\bibnamefont {Tokura}}, \bibinfo {author} {\bibfnamefont {Y.-S.}\ \bibnamefont {Shin}}, \bibinfo {author} {\bibfnamefont {T.}~\bibnamefont {Kubo}}, \bibinfo {author} {\bibfnamefont {K.}~\bibnamefont {Yoshida}}, \bibinfo {author} {\bibfnamefont {T.}~\bibnamefont {Taniyama}},\ and\ \bibinfo {author} {\bibfnamefont {S.}~\bibnamefont {Tarucha}},\ }\bibfield  {title} {\bibinfo {title} {{Electrically driven single-electron spin resonance in a slanting Zeeman field}},\ }\href {https://doi.org/10.1038/nphys1053} {\bibfield  {journal} {\bibinfo  {journal} {Nat. Phys.}\ }\textbf {\bibinfo {volume} {4}},\ \bibinfo {pages} {776} (\bibinfo {year} {2008})}\BibitemShut {NoStop}%
\bibitem [{\citenamefont {Kanai}\ \emph {et~al.}(2024)\citenamefont {Kanai}, \citenamefont {Jin},\ and\ \citenamefont {Guo}}]{Kanai2024-bo}%
  \BibitemOpen
  \bibfield  {author} {\bibinfo {author} {\bibfnamefont {T.}~\bibnamefont {Kanai}}, \bibinfo {author} {\bibfnamefont {D.}~\bibnamefont {Jin}},\ and\ \bibinfo {author} {\bibfnamefont {W.}~\bibnamefont {Guo}},\ }\bibfield  {title} {\bibinfo {title} {Single-electron qubits based on quantum ring states on solid neon surface},\ }\href {https://doi.org/10.1103/PhysRevLett.132.250603} {\bibfield  {journal} {\bibinfo  {journal} {Phys. Rev. Lett.}\ }\textbf {\bibinfo {volume} {132}},\ \bibinfo {pages} {250603} (\bibinfo {year} {2024})}\BibitemShut {NoStop}%
\bibitem [{\citenamefont {Goldman}\ \emph {et~al.}(2000)\citenamefont {Goldman}, \citenamefont {Ladd}, \citenamefont {Yamaguchi},\ and\ \citenamefont {Yamamoto}}]{Goldman}%
  \BibitemOpen
  \bibfield  {author} {\bibinfo {author} {\bibfnamefont {J.}~\bibnamefont {Goldman}}, \bibinfo {author} {\bibfnamefont {T.}~\bibnamefont {Ladd}}, \bibinfo {author} {\bibfnamefont {F.}~\bibnamefont {Yamaguchi}},\ and\ \bibinfo {author} {\bibfnamefont {Y.}~\bibnamefont {Yamamoto}},\ }\bibfield  {title} {\bibinfo {title} {{Magnet designs for a crystal-lattice quantum computer}},\ }\href {https://doi.org/10.1007/PL00021084} {\bibfield  {journal} {\bibinfo  {journal} {Applied Physics A}\ }\textbf {\bibinfo {volume} {71}},\ \bibinfo {pages} {11} (\bibinfo {year} {2000})}\BibitemShut {NoStop}%
\bibitem [{Note1()}]{Note1}%
  \BibitemOpen
  \bibinfo {note} {In analogy with Ref.~\protect \citenum {Benito2019-pi}, the decay due to quasi-static noise is assumed to follow a Gaussian form, $\exp (-(\gamma ^*t)^2)$.}\BibitemShut {Stop}%
\bibitem [{\citenamefont {Benito}\ \emph {et~al.}(2019{\natexlab{b}})\citenamefont {Benito}, \citenamefont {Croot}, \citenamefont {Adelsberger}, \citenamefont {Putz}, \citenamefont {Mi}, \citenamefont {Petta},\ and\ \citenamefont {Burkard}}]{Benito2019-pi}%
  \BibitemOpen
  \bibfield  {author} {\bibinfo {author} {\bibfnamefont {M.}~\bibnamefont {Benito}}, \bibinfo {author} {\bibfnamefont {X.}~\bibnamefont {Croot}}, \bibinfo {author} {\bibfnamefont {C.}~\bibnamefont {Adelsberger}}, \bibinfo {author} {\bibfnamefont {S.}~\bibnamefont {Putz}}, \bibinfo {author} {\bibfnamefont {X.}~\bibnamefont {Mi}}, \bibinfo {author} {\bibfnamefont {J.~R.}\ \bibnamefont {Petta}},\ and\ \bibinfo {author} {\bibfnamefont {G.}~\bibnamefont {Burkard}},\ }\bibfield  {title} {\bibinfo {title} {Electric-field control and noise protection of the flopping-mode spin qubit},\ }\href@noop {} {\bibfield  {journal} {\bibinfo  {journal} {Phys. Rev. B.}\ }\textbf {\bibinfo {volume} {100}},\ \bibinfo {pages} {125430} (\bibinfo {year} {2019}{\natexlab{b}})}\BibitemShut {NoStop}%
\bibitem [{\citenamefont {Kim}\ \emph {et~al.}(2022)\citenamefont {Kim}, \citenamefont {Yun}, \citenamefont {Jang}, \citenamefont {Jang}, \citenamefont {Park}, \citenamefont {Song}, \citenamefont {Cho}, \citenamefont {Sim}, \citenamefont {Sohn}, \citenamefont {Jung}, \citenamefont {Umansky},\ and\ \citenamefont {Kim}}]{Kim2022-jq}%
  \BibitemOpen
  \bibfield  {author} {\bibinfo {author} {\bibfnamefont {J.}~\bibnamefont {Kim}}, \bibinfo {author} {\bibfnamefont {J.}~\bibnamefont {Yun}}, \bibinfo {author} {\bibfnamefont {W.}~\bibnamefont {Jang}}, \bibinfo {author} {\bibfnamefont {H.}~\bibnamefont {Jang}}, \bibinfo {author} {\bibfnamefont {J.}~\bibnamefont {Park}}, \bibinfo {author} {\bibfnamefont {Y.}~\bibnamefont {Song}}, \bibinfo {author} {\bibfnamefont {M.-K.}\ \bibnamefont {Cho}}, \bibinfo {author} {\bibfnamefont {S.}~\bibnamefont {Sim}}, \bibinfo {author} {\bibfnamefont {H.}~\bibnamefont {Sohn}}, \bibinfo {author} {\bibfnamefont {H.}~\bibnamefont {Jung}}, \bibinfo {author} {\bibfnamefont {V.}~\bibnamefont {Umansky}},\ and\ \bibinfo {author} {\bibfnamefont {D.}~\bibnamefont {Kim}},\ }\bibfield  {title} {\bibinfo {title} {Approaching ideal visibility in singlet-triplet qubit operations using energy-selective tunneling-based hamiltonian estimation},\ }\href@noop {} {\bibfield  {journal} {\bibinfo  {journal} {Phys. Rev. Lett.}\ }\textbf {\bibinfo
  {volume} {129}},\ \bibinfo {pages} {040501} (\bibinfo {year} {2022})}\BibitemShut {NoStop}%
\bibitem [{\citenamefont {Berritta}\ \emph {et~al.}(2024)\citenamefont {Berritta}, \citenamefont {Rasmussen}, \citenamefont {Krzywda}, \citenamefont {van~der Heijden}, \citenamefont {Fedele}, \citenamefont {Fallahi}, \citenamefont {Gardner}, \citenamefont {Manfra}, \citenamefont {van Nieuwenburg}, \citenamefont {Danon}, \citenamefont {Chatterjee},\ and\ \citenamefont {Kuemmeth}}]{Berritta2024-bb}%
  \BibitemOpen
  \bibfield  {author} {\bibinfo {author} {\bibfnamefont {F.}~\bibnamefont {Berritta}}, \bibinfo {author} {\bibfnamefont {T.}~\bibnamefont {Rasmussen}}, \bibinfo {author} {\bibfnamefont {J.~A.}\ \bibnamefont {Krzywda}}, \bibinfo {author} {\bibfnamefont {J.}~\bibnamefont {van~der Heijden}}, \bibinfo {author} {\bibfnamefont {F.}~\bibnamefont {Fedele}}, \bibinfo {author} {\bibfnamefont {S.}~\bibnamefont {Fallahi}}, \bibinfo {author} {\bibfnamefont {G.~C.}\ \bibnamefont {Gardner}}, \bibinfo {author} {\bibfnamefont {M.~J.}\ \bibnamefont {Manfra}}, \bibinfo {author} {\bibfnamefont {E.}~\bibnamefont {van Nieuwenburg}}, \bibinfo {author} {\bibfnamefont {J.}~\bibnamefont {Danon}}, \bibinfo {author} {\bibfnamefont {A.}~\bibnamefont {Chatterjee}},\ and\ \bibinfo {author} {\bibfnamefont {F.}~\bibnamefont {Kuemmeth}},\ }\bibfield  {title} {\bibinfo {title} {Real-time two-axis control of a spin qubit},\ }\href@noop {} {\bibfield  {journal} {\bibinfo  {journal} {Nat. Commun.}\ }\textbf {\bibinfo {volume} {15}},\ \bibinfo
  {pages} {1676} (\bibinfo {year} {2024})}\BibitemShut {NoStop}%
\bibitem [{\citenamefont {Nakajima}\ \emph {et~al.}(2021)\citenamefont {Nakajima}, \citenamefont {Kojima}, \citenamefont {Uehara}, \citenamefont {Noiri}, \citenamefont {Takeda}, \citenamefont {Kobayashi},\ and\ \citenamefont {Tarucha}}]{Nakajima2021-tf}%
  \BibitemOpen
  \bibfield  {author} {\bibinfo {author} {\bibfnamefont {T.}~\bibnamefont {Nakajima}}, \bibinfo {author} {\bibfnamefont {Y.}~\bibnamefont {Kojima}}, \bibinfo {author} {\bibfnamefont {Y.}~\bibnamefont {Uehara}}, \bibinfo {author} {\bibfnamefont {A.}~\bibnamefont {Noiri}}, \bibinfo {author} {\bibfnamefont {K.}~\bibnamefont {Takeda}}, \bibinfo {author} {\bibfnamefont {T.}~\bibnamefont {Kobayashi}},\ and\ \bibinfo {author} {\bibfnamefont {S.}~\bibnamefont {Tarucha}},\ }\bibfield  {title} {\bibinfo {title} {Real-time feedback control of charge sensing for quantum dot qubits},\ }\href@noop {} {\bibfield  {journal} {\bibinfo  {journal} {Phys. Rev. Appl.}\ }\textbf {\bibinfo {volume} {15}} (\bibinfo {year} {2021})}\BibitemShut {NoStop}%
\bibitem [{\citenamefont {Nakajima}\ \emph {et~al.}(2020)\citenamefont {Nakajima}, \citenamefont {Noiri}, \citenamefont {Kawasaki}, \citenamefont {Yoneda}, \citenamefont {Stano}, \citenamefont {Amaha}, \citenamefont {Otsuka}, \citenamefont {Takeda}, \citenamefont {Delbecq}, \citenamefont {Allison}, \citenamefont {Ludwig}, \citenamefont {Wieck}, \citenamefont {Loss},\ and\ \citenamefont {Tarucha}}]{Nakajima2020-hr}%
  \BibitemOpen
  \bibfield  {author} {\bibinfo {author} {\bibfnamefont {T.}~\bibnamefont {Nakajima}}, \bibinfo {author} {\bibfnamefont {A.}~\bibnamefont {Noiri}}, \bibinfo {author} {\bibfnamefont {K.}~\bibnamefont {Kawasaki}}, \bibinfo {author} {\bibfnamefont {J.}~\bibnamefont {Yoneda}}, \bibinfo {author} {\bibfnamefont {P.}~\bibnamefont {Stano}}, \bibinfo {author} {\bibfnamefont {S.}~\bibnamefont {Amaha}}, \bibinfo {author} {\bibfnamefont {T.}~\bibnamefont {Otsuka}}, \bibinfo {author} {\bibfnamefont {K.}~\bibnamefont {Takeda}}, \bibinfo {author} {\bibfnamefont {M.~R.}\ \bibnamefont {Delbecq}}, \bibinfo {author} {\bibfnamefont {G.}~\bibnamefont {Allison}}, \bibinfo {author} {\bibfnamefont {A.}~\bibnamefont {Ludwig}}, \bibinfo {author} {\bibfnamefont {A.~D.}\ \bibnamefont {Wieck}}, \bibinfo {author} {\bibfnamefont {D.}~\bibnamefont {Loss}},\ and\ \bibinfo {author} {\bibfnamefont {S.}~\bibnamefont {Tarucha}},\ }\bibfield  {title} {\bibinfo {title} {Coherence of a driven electron spin qubit actively decoupled from quasistatic
  noise},\ }\href@noop {} {\bibfield  {journal} {\bibinfo  {journal} {Phys. Rev. X.}\ }\textbf {\bibinfo {volume} {10}} (\bibinfo {year} {2020})}\BibitemShut {NoStop}%
\bibitem [{\citenamefont {Shulman}\ \emph {et~al.}(2014)\citenamefont {Shulman}, \citenamefont {Harvey}, \citenamefont {Nichol}, \citenamefont {Bartlett}, \citenamefont {Doherty}, \citenamefont {Umansky},\ and\ \citenamefont {Yacoby}}]{Shulman2014-br}%
  \BibitemOpen
  \bibfield  {author} {\bibinfo {author} {\bibfnamefont {M.~D.}\ \bibnamefont {Shulman}}, \bibinfo {author} {\bibfnamefont {S.~P.}\ \bibnamefont {Harvey}}, \bibinfo {author} {\bibfnamefont {J.~M.}\ \bibnamefont {Nichol}}, \bibinfo {author} {\bibfnamefont {S.~D.}\ \bibnamefont {Bartlett}}, \bibinfo {author} {\bibfnamefont {A.~C.}\ \bibnamefont {Doherty}}, \bibinfo {author} {\bibfnamefont {V.}~\bibnamefont {Umansky}},\ and\ \bibinfo {author} {\bibfnamefont {A.}~\bibnamefont {Yacoby}},\ }\bibfield  {title} {\bibinfo {title} {Suppressing qubit dephasing using real-time hamiltonian estimation},\ }\href@noop {} {\bibfield  {journal} {\bibinfo  {journal} {Nat. Commun.}\ }\textbf {\bibinfo {volume} {5}},\ \bibinfo {pages} {5156} (\bibinfo {year} {2014})}\BibitemShut {NoStop}%
\bibitem [{\citenamefont {Park}\ \emph {et~al.}(2025)\citenamefont {Park}, \citenamefont {Jang}, \citenamefont {Sohn}, \citenamefont {Yun}, \citenamefont {Song}, \citenamefont {Kang}, \citenamefont {Stehouwer}, \citenamefont {Esposti}, \citenamefont {Scappucci},\ and\ \citenamefont {Kim}}]{Park2025-cy}%
  \BibitemOpen
  \bibfield  {author} {\bibinfo {author} {\bibfnamefont {J.}~\bibnamefont {Park}}, \bibinfo {author} {\bibfnamefont {H.}~\bibnamefont {Jang}}, \bibinfo {author} {\bibfnamefont {H.}~\bibnamefont {Sohn}}, \bibinfo {author} {\bibfnamefont {J.}~\bibnamefont {Yun}}, \bibinfo {author} {\bibfnamefont {Y.}~\bibnamefont {Song}}, \bibinfo {author} {\bibfnamefont {B.}~\bibnamefont {Kang}}, \bibinfo {author} {\bibfnamefont {L.~E.~A.}\ \bibnamefont {Stehouwer}}, \bibinfo {author} {\bibfnamefont {D.~D.}\ \bibnamefont {Esposti}}, \bibinfo {author} {\bibfnamefont {G.}~\bibnamefont {Scappucci}},\ and\ \bibinfo {author} {\bibfnamefont {D.}~\bibnamefont {Kim}},\ }\bibfield  {title} {\bibinfo {title} {Passive and active suppression of transduced noise in silicon spin qubits},\ }\href@noop {} {\bibfield  {journal} {\bibinfo  {journal} {Nat. Commun.}\ }\textbf {\bibinfo {volume} {16}},\ \bibinfo {pages} {78} (\bibinfo {year} {2025})}\BibitemShut {NoStop}%
\bibitem [{\citenamefont {Burkard}\ \emph {et~al.}(2020)\citenamefont {Burkard}, \citenamefont {Gullans}, \citenamefont {Mi},\ and\ \citenamefont {Petta}}]{Burkard2020-gy}%
  \BibitemOpen
  \bibfield  {author} {\bibinfo {author} {\bibfnamefont {G.}~\bibnamefont {Burkard}}, \bibinfo {author} {\bibfnamefont {M.~J.}\ \bibnamefont {Gullans}}, \bibinfo {author} {\bibfnamefont {X.}~\bibnamefont {Mi}},\ and\ \bibinfo {author} {\bibfnamefont {J.~R.}\ \bibnamefont {Petta}},\ }\bibfield  {title} {\bibinfo {title} {Superconductor--semiconductor hybrid-circuit quantum electrodynamics},\ }\href@noop {} {\bibfield  {journal} {\bibinfo  {journal} {Nat. Rev. Phys.}\ }\textbf {\bibinfo {volume} {2}},\ \bibinfo {pages} {129} (\bibinfo {year} {2020})}\BibitemShut {NoStop}%
\bibitem [{\citenamefont {Warren}\ \emph {et~al.}(2019)\citenamefont {Warren}, \citenamefont {Barnes},\ and\ \citenamefont {Economou}}]{Warren2019-mn}%
  \BibitemOpen
  \bibfield  {author} {\bibinfo {author} {\bibfnamefont {A.}~\bibnamefont {Warren}}, \bibinfo {author} {\bibfnamefont {E.}~\bibnamefont {Barnes}},\ and\ \bibinfo {author} {\bibfnamefont {S.~E.}\ \bibnamefont {Economou}},\ }\bibfield  {title} {\bibinfo {title} {Long-distance entangling gates between quantum dot spins mediated by a superconducting resonator},\ }\href@noop {} {\bibfield  {journal} {\bibinfo  {journal} {Phys. Rev. B.}\ }\textbf {\bibinfo {volume} {100}} (\bibinfo {year} {2019})}\BibitemShut {NoStop}%
\bibitem [{\citenamefont {Ball}\ \emph {et~al.}(2016)\citenamefont {Ball}, \citenamefont {Oliver},\ and\ \citenamefont {Biercuk}}]{Ball2016-zn}%
  \BibitemOpen
  \bibfield  {author} {\bibinfo {author} {\bibfnamefont {H.}~\bibnamefont {Ball}}, \bibinfo {author} {\bibfnamefont {W.}~\bibnamefont {Oliver}},\ and\ \bibinfo {author} {\bibfnamefont {M.}~\bibnamefont {Biercuk}},\ }\bibfield  {title} {\bibinfo {title} {The role of master clock stability in quantum information processing},\ }\href {https://doi.org/10.1038/npjqi.2016.33} {\bibfield  {journal} {\bibinfo  {journal} {npj Quantum Information}\ }\textbf {\bibinfo {volume} {2}},\ \bibinfo {pages} {16033} (\bibinfo {year} {2016})}\BibitemShut {NoStop}%
\bibitem [{\citenamefont {Mi}\ \emph {et~al.}(2017{\natexlab{b}})\citenamefont {Mi}, \citenamefont {Cady}, \citenamefont {Zajac}, \citenamefont {Stehlik}, \citenamefont {Edge},\ and\ \citenamefont {Petta}}]{Mi2017-go}%
  \BibitemOpen
  \bibfield  {author} {\bibinfo {author} {\bibfnamefont {X.}~\bibnamefont {Mi}}, \bibinfo {author} {\bibfnamefont {J.~V.}\ \bibnamefont {Cady}}, \bibinfo {author} {\bibfnamefont {D.~M.}\ \bibnamefont {Zajac}}, \bibinfo {author} {\bibfnamefont {J.}~\bibnamefont {Stehlik}}, \bibinfo {author} {\bibfnamefont {L.~F.}\ \bibnamefont {Edge}},\ and\ \bibinfo {author} {\bibfnamefont {J.~R.}\ \bibnamefont {Petta}},\ }\bibfield  {title} {\bibinfo {title} {Circuit quantum electrodynamics architecture for gate-defined quantum dots in silicon},\ }\href@noop {} {\bibfield  {journal} {\bibinfo  {journal} {Applied Physics Letters}\ }\textbf {\bibinfo {volume} {110}},\ \bibinfo {pages} {043502} (\bibinfo {year} {2017}{\natexlab{b}})}\BibitemShut {NoStop}%
\bibitem [{\citenamefont {Harvey-Collard}\ \emph {et~al.}(2020)\citenamefont {Harvey-Collard}, \citenamefont {Zheng}, \citenamefont {Dijkema}, \citenamefont {Samkharadze}, \citenamefont {Sammak}, \citenamefont {Scappucci},\ and\ \citenamefont {Vandersypen}}]{Harvey-Collard2020-dt}%
  \BibitemOpen
  \bibfield  {author} {\bibinfo {author} {\bibfnamefont {P.}~\bibnamefont {Harvey-Collard}}, \bibinfo {author} {\bibfnamefont {G.}~\bibnamefont {Zheng}}, \bibinfo {author} {\bibfnamefont {J.}~\bibnamefont {Dijkema}}, \bibinfo {author} {\bibfnamefont {N.}~\bibnamefont {Samkharadze}}, \bibinfo {author} {\bibfnamefont {A.}~\bibnamefont {Sammak}}, \bibinfo {author} {\bibfnamefont {G.}~\bibnamefont {Scappucci}},\ and\ \bibinfo {author} {\bibfnamefont {L.~M.~K.}\ \bibnamefont {Vandersypen}},\ }\bibfield  {title} {\bibinfo {title} {On-chip microwave filters for high-impedance resonators with gate-defined quantum dots},\ }\href@noop {} {\bibfield  {journal} {\bibinfo  {journal} {Physical Review Applied}\ }\textbf {\bibinfo {volume} {14}},\ \bibinfo {pages} {034025} (\bibinfo {year} {2020})}\BibitemShut {NoStop}%
\bibitem [{\citenamefont {Gao}(2008)}]{gao2008physics}%
  \BibitemOpen
  \bibfield  {author} {\bibinfo {author} {\bibfnamefont {J.}~\bibnamefont {Gao}},\ }{\selectlanguage {en}\emph {\bibinfo {title} {The physics of superconducting microwave resonators}}},\ \href@noop {} {Ph.D. thesis},\ \bibinfo  {school} {California Institute of Technology} (\bibinfo {year} {2008})\BibitemShut {NoStop}%
\bibitem [{\citenamefont {Andrei}(1997)}]{Andrei1997Two-DimensionalSubstrates}%
  \BibitemOpen
  \bibfield  {author} {\bibinfo {author} {\bibfnamefont {E.~Y.}\ \bibnamefont {Andrei}},\ }\href@noop {} {\emph {\bibinfo {title} {Two-Dimensional Electron Systems : on Helium and other Cryogenic Substrates}}}\ (\bibinfo  {publisher} {Springer Netherlands},\ \bibinfo {year} {1997})\BibitemShut {NoStop}%
\bibitem [{Note2()}]{Note2}%
  \BibitemOpen
  \bibinfo {note} {Note that we use the convention \(\exp (i \omega t)\) for the time dependence of fields. While the physics community commonly adopts \(\exp (-i \omega t)\), the engineering convention \(\exp (i \omega t)\) is used here for consistency with COMSOL.}\BibitemShut {Stop}%
\bibitem [{\citenamefont {Pozar}(1998)}]{Pozar1998}%
  \BibitemOpen
  \bibfield  {author} {\bibinfo {author} {\bibfnamefont {D.~M.}\ \bibnamefont {Pozar}},\ }\href@noop {} {\emph {\bibinfo {title} {Microwave engineering}}},\ \bibinfo {edition} {4th}\ ed.\ (\bibinfo  {publisher} {John Wiley \& Sons Inc.,},\ \bibinfo {address} {New York},\ \bibinfo {year} {1998})\BibitemShut {NoStop}%
\bibitem [{\citenamefont {Gurioli}\ \emph {et~al.}(1992)\citenamefont {Gurioli}, \citenamefont {Martinez-Pastor}, \citenamefont {Colocci}, \citenamefont {Deparis}, \citenamefont {Chastaingt},\ and\ \citenamefont {Massies}}]{Gurioli1992-ms}%
  \BibitemOpen
  \bibfield  {author} {\bibinfo {author} {\bibfnamefont {M.}~\bibnamefont {Gurioli}}, \bibinfo {author} {\bibfnamefont {J.}~\bibnamefont {Martinez-Pastor}}, \bibinfo {author} {\bibfnamefont {M.}~\bibnamefont {Colocci}}, \bibinfo {author} {\bibfnamefont {C.}~\bibnamefont {Deparis}}, \bibinfo {author} {\bibfnamefont {B.}~\bibnamefont {Chastaingt}},\ and\ \bibinfo {author} {\bibfnamefont {J.}~\bibnamefont {Massies}},\ }\bibfield  {title} {\bibinfo {title} {Thermal escape of carriers out of gaas/alxga1-xas quantum-well structures},\ }\href@noop {} {\bibfield  {journal} {\bibinfo  {journal} {Phys. Rev. B}\ }\textbf {\bibinfo {volume} {46}},\ \bibinfo {pages} {6922} (\bibinfo {year} {1992})}\BibitemShut {NoStop}%
\bibitem [{\citenamefont {Swank}\ and\ \citenamefont {Brown}(1963)}]{Swank1963-ap}%
  \BibitemOpen
  \bibfield  {author} {\bibinfo {author} {\bibfnamefont {R.~K.}\ \bibnamefont {Swank}}\ and\ \bibinfo {author} {\bibfnamefont {F.~C.}\ \bibnamefont {Brown}},\ }\bibfield  {title} {\bibinfo {title} {Lifetime of the excited f center},\ }\href@noop {} {\bibfield  {journal} {\bibinfo  {journal} {Phys. Rev.}\ }\textbf {\bibinfo {volume} {130}},\ \bibinfo {pages} {34} (\bibinfo {year} {1963})}\BibitemShut {NoStop}%
\bibitem [{\citenamefont {Gao}\ \emph {et~al.}(2008)\citenamefont {Gao}, \citenamefont {Daal}, \citenamefont {Vayonakis}, \citenamefont {Kumar}, \citenamefont {Zmuidzinas}, \citenamefont {Sadoulet}, \citenamefont {Mazin}, \citenamefont {Day},\ and\ \citenamefont {Leduc}}]{Gao2008-ox}%
  \BibitemOpen
  \bibfield  {author} {\bibinfo {author} {\bibfnamefont {J.}~\bibnamefont {Gao}}, \bibinfo {author} {\bibfnamefont {M.}~\bibnamefont {Daal}}, \bibinfo {author} {\bibfnamefont {A.}~\bibnamefont {Vayonakis}}, \bibinfo {author} {\bibfnamefont {S.}~\bibnamefont {Kumar}}, \bibinfo {author} {\bibfnamefont {J.}~\bibnamefont {Zmuidzinas}}, \bibinfo {author} {\bibfnamefont {B.}~\bibnamefont {Sadoulet}}, \bibinfo {author} {\bibfnamefont {B.~A.}\ \bibnamefont {Mazin}}, \bibinfo {author} {\bibfnamefont {P.~K.}\ \bibnamefont {Day}},\ and\ \bibinfo {author} {\bibfnamefont {H.~G.}\ \bibnamefont {Leduc}},\ }\bibfield  {title} {\bibinfo {title} {Experimental evidence for a surface distribution of two-level systems in superconducting lithographed microwave resonators},\ }\href@noop {} {\bibfield  {journal} {\bibinfo  {journal} {Appl. Phys. Lett.}\ }\textbf {\bibinfo {volume} {92}},\ \bibinfo {pages} {152505} (\bibinfo {year} {2008})}\BibitemShut {NoStop}%
\bibitem [{\citenamefont {M{\"u}ller}\ \emph {et~al.}(2022)\citenamefont {M{\"u}ller}, \citenamefont {Luschmann}, \citenamefont {Faltermeier}, \citenamefont {Weichselbaumer}, \citenamefont {Koch}, \citenamefont {Huber}, \citenamefont {Schumacher}, \citenamefont {Ubbelohde}, \citenamefont {Reifert}, \citenamefont {Scheller} \emph {et~al.}}]{muller2022magnetic}%
  \BibitemOpen
  \bibfield  {author} {\bibinfo {author} {\bibfnamefont {M.}~\bibnamefont {M{\"u}ller}}, \bibinfo {author} {\bibfnamefont {T.}~\bibnamefont {Luschmann}}, \bibinfo {author} {\bibfnamefont {A.}~\bibnamefont {Faltermeier}}, \bibinfo {author} {\bibfnamefont {S.}~\bibnamefont {Weichselbaumer}}, \bibinfo {author} {\bibfnamefont {L.}~\bibnamefont {Koch}}, \bibinfo {author} {\bibfnamefont {G.~B.}\ \bibnamefont {Huber}}, \bibinfo {author} {\bibfnamefont {H.~W.}\ \bibnamefont {Schumacher}}, \bibinfo {author} {\bibfnamefont {N.}~\bibnamefont {Ubbelohde}}, \bibinfo {author} {\bibfnamefont {D.}~\bibnamefont {Reifert}}, \bibinfo {author} {\bibfnamefont {T.}~\bibnamefont {Scheller}}, \emph {et~al.},\ }\bibfield  {title} {\bibinfo {title} {Magnetic field robust high quality factor nbtin superconducting microwave resonators},\ }\href@noop {} {\bibfield  {journal} {\bibinfo  {journal} {Materials for Quantum Technology}\ }\textbf {\bibinfo {volume} {2}},\ \bibinfo {pages} {015002} (\bibinfo {year} {2022})}\BibitemShut {NoStop}%
\bibitem [{\citenamefont {Bruno}\ \emph {et~al.}(2015)\citenamefont {Bruno}, \citenamefont {De~Lange}, \citenamefont {Asaad}, \citenamefont {Van Der~Enden}, \citenamefont {Langford},\ and\ \citenamefont {DiCarlo}}]{bruno2015reducing}%
  \BibitemOpen
  \bibfield  {author} {\bibinfo {author} {\bibfnamefont {A.}~\bibnamefont {Bruno}}, \bibinfo {author} {\bibfnamefont {G.}~\bibnamefont {De~Lange}}, \bibinfo {author} {\bibfnamefont {S.}~\bibnamefont {Asaad}}, \bibinfo {author} {\bibfnamefont {K.}~\bibnamefont {Van Der~Enden}}, \bibinfo {author} {\bibfnamefont {N.}~\bibnamefont {Langford}},\ and\ \bibinfo {author} {\bibfnamefont {L.}~\bibnamefont {DiCarlo}},\ }\bibfield  {title} {\bibinfo {title} {Reducing intrinsic loss in superconducting resonators by surface treatment and deep etching of silicon substrates},\ }\href@noop {} {\bibfield  {journal} {\bibinfo  {journal} {Applied Physics Letters}\ }\textbf {\bibinfo {volume} {106}} (\bibinfo {year} {2015})}\BibitemShut {NoStop}%
\bibitem [{\citenamefont {Barends}\ \emph {et~al.}(2010)\citenamefont {Barends}, \citenamefont {Vercruyssen}, \citenamefont {Endo}, \citenamefont {De~Visser}, \citenamefont {Zijlstra}, \citenamefont {Klapwijk}, \citenamefont {Diener}, \citenamefont {Yates},\ and\ \citenamefont {Baselmans}}]{barends2010minimal}%
  \BibitemOpen
  \bibfield  {author} {\bibinfo {author} {\bibfnamefont {R.}~\bibnamefont {Barends}}, \bibinfo {author} {\bibfnamefont {N.}~\bibnamefont {Vercruyssen}}, \bibinfo {author} {\bibfnamefont {A.}~\bibnamefont {Endo}}, \bibinfo {author} {\bibfnamefont {P.}~\bibnamefont {De~Visser}}, \bibinfo {author} {\bibfnamefont {T.}~\bibnamefont {Zijlstra}}, \bibinfo {author} {\bibfnamefont {T.}~\bibnamefont {Klapwijk}}, \bibinfo {author} {\bibfnamefont {P.}~\bibnamefont {Diener}}, \bibinfo {author} {\bibfnamefont {S.}~\bibnamefont {Yates}},\ and\ \bibinfo {author} {\bibfnamefont {J.}~\bibnamefont {Baselmans}},\ }\bibfield  {title} {\bibinfo {title} {Minimal resonator loss for circuit quantum electrodynamics},\ }\href@noop {} {\bibfield  {journal} {\bibinfo  {journal} {Applied Physics Letters}\ }\textbf {\bibinfo {volume} {97}} (\bibinfo {year} {2010})}\BibitemShut {NoStop}%
\bibitem [{\citenamefont {Lane}\ \emph {et~al.}(2020)\citenamefont {Lane}, \citenamefont {Tan}, \citenamefont {Beysengulov}, \citenamefont {Nasyedkin}, \citenamefont {Brook}, \citenamefont {Zhang}, \citenamefont {Stefanski}, \citenamefont {Byeon}, \citenamefont {Murch},\ and\ \citenamefont {Pollanen}}]{Lane2020-pc}%
  \BibitemOpen
  \bibfield  {author} {\bibinfo {author} {\bibfnamefont {J.~R.}\ \bibnamefont {Lane}}, \bibinfo {author} {\bibfnamefont {D.}~\bibnamefont {Tan}}, \bibinfo {author} {\bibfnamefont {N.~R.}\ \bibnamefont {Beysengulov}}, \bibinfo {author} {\bibfnamefont {K.}~\bibnamefont {Nasyedkin}}, \bibinfo {author} {\bibfnamefont {E.}~\bibnamefont {Brook}}, \bibinfo {author} {\bibfnamefont {L.}~\bibnamefont {Zhang}}, \bibinfo {author} {\bibfnamefont {T.}~\bibnamefont {Stefanski}}, \bibinfo {author} {\bibfnamefont {H.}~\bibnamefont {Byeon}}, \bibinfo {author} {\bibfnamefont {K.~W.}\ \bibnamefont {Murch}},\ and\ \bibinfo {author} {\bibfnamefont {J.}~\bibnamefont {Pollanen}},\ }\bibfield  {title} {\bibinfo {title} {Integrating superfluids with superconducting qubit systems},\ }\href@noop {} {\bibfield  {journal} {\bibinfo  {journal} {Phys. Rev. A}\ }\textbf {\bibinfo {volume} {101}},\ \bibinfo {pages} {012336} (\bibinfo {year} {2020})}\BibitemShut {NoStop}%
\end{thebibliography}%

\end{document}